\documentstyle[epsf]{l-aa}
\begin{document}
\thesaurus{06(02.18.7, 03.13.4, 08.14.1, 08.19.4, 02.05.1)}

\title{Neutrino transport in type II supernovae:$\ $\\
Boltzmann solver vs.~Monte Carlo method}
\author{Shoichi Yamada\inst{1,}\inst{2}
\thanks{e-mail: shoichi@MPA-Garching.MPG.DE} 
\and Hans-Thomas Janka\inst{1}\thanks{e-mail: thj@MPA-Garching.MPG.DE} 
\and Hideyuki Suzuki\inst{1,}\inst{3}\thanks{e-mail: hideyuki.suzuki@kek.jp}
}
\offprints{S.~Yamada}
\institute{Max-Planck-Institut f\"{u}r Astrophysik,
Karl-Schwarzschild-Str.~1, Postfach 1523, D--85740 Garching, Germany
\and
Department of Physics, Graduate School of Science, 
The University of Tokyo, 7--3--1, Hongo, Bunkyo-ku, Tokyo 113, Japan
\and
High Energy Accelerator Research Organization (KEK),
Oho, Tsukuba, Ibaraki 305-0801, Japan}
\date{Received; accepted}
\maketitle

\begin{abstract}

We have coded a Boltzmann solver based on a finite difference scheme 
(S$_N$ method) aiming at calculations of neutrino transport in type II
supernovae. Close comparison between the Boltzmann solver and a Monte
Carlo transport code has been made for realistic atmospheres of post 
bounce core models under the assumption of a static background.
We have also investigated in detail the dependence of the results
on the numbers of radial, angular, and energy grid points
and the way to discretize the spatial advection term which is used in the 
Boltzmann solver. A general relativistic calculation has been done for
one of the models. We find overall good 
agreement between the two methods. This gives credibility to both
methods which are based on completely different formulations.
In particular, the number and energy fluxes and the mean energies
of the neutrinos show remarkably good agreement, because these 
quantities are
determined in a region where the angular distribution of the
neutrinos is nearly isotropic and they are essentially frozen in
later on. On the other hand, because of a relatively small number of
angular grid points (which is inevitable due to limitations
of the computation time) the Boltzmann solver tends to underestimate
the flux factor and the Eddington factor outside the (mean)
``neutrinosphere'' where the angular distribution of the neutrinos
becomes highly anisotropic. As a result, the neutrino number density 
is overestimated in this region. This fact suggests that one has to
be cautious in applying the Boltzmann solver to a calculation of the
neutrino heating in the hot-bubble region because it might tend to
overestimate the local energy deposition rate. A comparison shows that
this trend is opposite to the results obtained with a multi-group 
flux-limited diffusion approximation of neutrino transport,
employing three different flux limiters, all of which lead to an 
underestimation of the hot-bubble heating. The accuracy of the
Boltzmann solver can be considerably improved by using a variable 
angular mesh to increase the angular resolution in the 
semi-transparent regime. 
\end{abstract}

\section{Introduction}
Neutrino energy transfer to the matter adjacent to the nascent
neutron star is supposed to trigger the explosion
of a massive star ($M \ga 8\,M_\odot$) as a type II supernova.
Since the energy released in neutrinos by the collapsed stellar 
iron core is more than 100 times larger 
than the kinetic energy of the explosion, only a small fraction of 
the neutrino energy is sufficient to expel the mantle and envelope
of the progenitor star. Numerical simulations have demonstrated the 
viability of this neutrino-driven explosion mechanism (Wilson~\cite{wi82};
Bethe \& Wilson~\cite{bw85}; Wilson~et~al.~\cite{wm86}) but the
explosions turned out to be sensitive to the size of the neutrino
luminosities and the neutrino spectra (Janka~\cite{ja93}; 
Janka \& M\"uller~\cite{jm93}; Burrows \& Goshy~\cite{bg93}) 
both of which determine the power of
the neutrino energy transfer to the matter outside the average
neutrinosphere. The rate of energy deposition per nucleon via
the dominant processes of electron neutrino absorption on neutrons
and electron antineutrino absorption on protons is given by:
\begin{equation}
Q_{\nu}^+\,\approx\, 110\cdot {L_{\nu,52}\langle\epsilon_{\nu,15}^2\rangle
\over r_7^2 \,\, \left\langle \mu \right\rangle}
\cdot\left\{\matrix{Y_n\cr Y_p \cr}\right \}\quad
\left\lbrack {{\rm MeV}\over {\rm s}\cdot N}\right\rbrack \; .
\label{eq-1}
\end{equation}
Here $Y_n = n_n/n_b$ and $Y_p = n_p/n_b$ are the number fractions of 
free neutrons and protons, respectively; the normalization with the 
baryon density $n_b$ indicates that the rate per baryon is calculated
in Eq.~(\ref{eq-1}). $L_{\nu,52}$ denotes the luminosity of either
$\nu_e$ or $\bar\nu_e$ in
units of $10^{52}\,{\rm erg/s}$ and $r_7$ is the radial position in 
$10^7\,{\rm cm}$. The average of the squared neutrino energy,
$\langle\epsilon_{\nu,15}^2\rangle$, is measured in units of 
$(15\,{\rm MeV})^2$ and enters through the energy dependence
of the neutrino and antineutrino absorption cross sections.
$\left\langle \mu \right\rangle$ is the angular dilution factor
of the neutrino radiation field (the ``flux factor'', which is equal to the mean
value of the cosine of the angle of neutrino propagation relative to the
radial direction) which varies between values much less than unity deep
inside the protoneutron star atmosphere, about 0.25 around the neutrinosphere,
and 1 for radially streaming neutrinos very far out. The factor
$\left\langle \mu \right\rangle$ determines the
local neutrino energy density according to the relation 
$E_{\nu} = L_{\nu}/(4\pi r^2 c \left\langle \mu \right\rangle)$ and
thus enters the heating rate of Eq.~(\ref{eq-1}). Only far away
from the neutrino emitting star, $\left\langle \mu \right\rangle\to 1$, 
and $E_{\nu}$ dilutes like $r^{-2}$.

Although it was found in two-dimensional simulations that convective 
instabilities in the neutrino-heating
region can help the explosion (Herant~et~al.~\cite{he94};
Janka \& M\"uller~\cite{jm93}, \cite{jm96}; 
Burrows~et~al.~\cite{bu95}; Miller~et~al.~\cite{mw93};
Shimizu~et~al.~\cite{sh94}) by the exchange of hot gas from the heating
layer with cold gas from the postshock region, the strength of this
convective overturn and its importance for the explosion is still a
matter of debate (Janka \& M\"uller~\cite{jm96}, 
Mezzacappa~et~al.~\cite{me98}, Lichtenstadt et al.~\cite{li98}).
In addition, it turns out that the development of an explosion
remains sensitive to the neutrino luminosities and the mean spectral 
energies even if convective overturn lowers the required threshold
values. This is the case because convective instabilities can develop 
sufficiently quickly only when the heating is fast and an unstable 
stratification builds up more quickly than the heated matter is
advected from the postshock region through the gain radius (which is the 
radius separating neutrino cooling inside from neutrino heating 
outside) down onto the neutron star surface
(Janka \& M\"uller~\cite{jm96}, Janka \& Keil~\cite{jk98}).

``Robust'' neutrino-driven explosions might
therefore require larger accretion luminosities (to be precise: a 
larger value of the product $L_{\nu}\langle\epsilon_{\nu}^2\rangle$
in Eq.~(\ref{eq-1})) during the early post-bounce phase, or 
might call for enhanced neutrino emission from the core.
The latter could be caused, for example, 
by convective neutrino transport within the nascent neutron star
(Burrows~\cite{bu87}; Mayle \& Wilson~\cite{mw88}; 
Wilson \& Mayle~\cite{wm88}, \cite{wm93};
Keil~et~al.~\cite{ke96}) or, alternatively, by a suppression of the neutrino 
opacities at nuclear densities through nucleon correlations 
(Sawyer~\cite{sa89}; Horowitz \& Wehrberger~\cite{ho91}; 
Raffelt \& Seckel~\cite{ra95}; Raffelt~et~al.~\cite{ra96}; 
Keil~et~al.~\cite{kj95}; 
Janka~et~al.~\cite{jk96}; Burrows \& Sawyer~\cite{bs98a},\cite{bs98b}; 
Reddy~et~al.~\cite{rp98a}), 
nucleon recoil and blocking (Schinder~\cite{sc90}) and/or nuclear interaction
effects in the neutrino-nucleon interactions (Prakash~et~al.~\cite{pl97}; 
Reddy~et~al.~\cite{rp97}, \cite{rp98b}), 
all of which have to date not been taken into account fully 
self-consistently in supernova simulations. The diffusive propagation 
of neutrinos out from the very opaque inner core is determined by the
value of the diffusion constant and thus sensitive to these effects.

Most of the current numerical treatments of neutrino transport, however,
are deficient not only concerning their description of the extremely complex 
neutrino interactions in the dense nuclear plasma but also concerning 
their handling of the transition from diffusion to free streaming.
While the core flux is fixed in the diffusive regime, the accretion
luminosity as well as the spectra of the emitted neutrinos depend
on the transport in the semitransparent layers around the sphere of
last scattering. Since neutrino-matter interactions are strongly
dependent on the neutrino energy, neutrinos with different energies
interact with largely different rates and 
decouple in layers with different densities and temperatures. The
spectral shape of the emergent neutrino flux is therefore different 
from the thermal spectrum at any particular point in the atmosphere.
Even more, through the factor $\left\langle \mu \right\rangle$ 
in the denominator of Eq.~(\ref{eq-1}) 
the energy deposition rate depends on the angular distribution 
of the neutrinos in the heating region. A quantitatively reliable
description of these aspects requires the use of sophisticated
transport algorithms which solve the Boltzmann equation instead
of approximate methods like flux-limited diffusion techniques 
(Janka~\cite{ja91a}, \cite{ja92}; 
Mezzacappa \& Bruenn~\cite{me93a},\cite{me93b},\cite{me93c};
Messer~et~al.~\cite{mm98}). 
The detection of electron antineutrinos from SN~1987A in the
Kamiokande~II (Hirata~et~al.~\cite{hi87}) and IMB laboratories 
(Bionta~et~al.~\cite{bi87})
and the construction of new, even larger neutrino experiments for
future supernova neutrino measurements have raised additional 
interest in accurate predictions of the detectable neutrino signals
from type II supernovae.

Neutrino transport in core collapse supernovae is a very complex problem 
and difficult to
treat accurately even in the spherically symmetric case. Some of
the major difficulties arise from the strong energy dependence of the
neutrino interactions, the non-conservative and anisotropic nature of the
scattering processes such as neutrino-electron scattering, the non-linearity
of the reaction kernels through neutrino Fermi blocking, and the need
to couple neutrino and antineutrino transport for the neutrino-pair
reactions. Therefore various simplifications and approximations have
been employed in numerical simulations of supernova explosions and
neutron star formation. The so far most widely used approximation
with a high degree of sophistication is the (multi-energy-group)
flux-limited diffusion (MGFLD) (Bowers \& Wilson~\cite{bw82}, 
Bruenn~\cite{br85}, Myra~et~al.~\cite{my87}, Suzuki~\cite{su90},
Lichtenstadt et al.~\cite{li98}) where a flux-limiting parameter is
employed in the formulation of the neutrino flux to ensure a smooth 
interpolation between the diffusion regime (where the neutrinos are 
essentially isotropic) and the free streaming regime (where the neutrinos
move radially outward). Although the limits are accurately reproduced,
there is no guarantee that the intermediate regime is properly
treated. Since in a quasi-stationary situation (e.g., for the cooling 
protoneutron star) the flux and the mean energy of the emitted
neutrinos are determined in the diffusion regime, little change of
these is found when the flux-limiter is varied (Suzuki~\cite{su90}) or the
transport equation is directly solved, e.g., by Monte Carlo calculations
(Janka~\cite{ja91a}). This, however, is not true when the spectral form is
considered, because the spectra are shaped in the semitransparent
surface-near layers. Moreover, significant differences are also 
expected for problems 
where the local angular distribution is important in the region between 
the diffusion and free streaming limits. Due to the factor
$\left\langle \mu \right\rangle$ appearing
in Eq.~(\ref{eq-1}) the hot-bubble heating is such a problem, 
neutrino-antineutrino pair annihilation is another problem of this
kind. In fact, Monte Carlo simulations 
(Janka \& Hillebrandt~\cite{jh89a},\cite{jh89b};
Janka~\cite{ja91a}; Janka~et~al.~\cite{jd92}) have shown
that all flux-limiters overestimate the anisotropy of the radiation
field outside the neutrinosphere, i.e., $\left\langle \mu
\right\rangle = 1$ is enforced too rapidly 
(see also Cernohorsky \& Bludman~\cite{cb94}).
This leads to an underestimation of the neutrino heating in the 
hot-bubble region between neutrinosphere and supernova shock
(Eq.~(\ref{eq-1})), and sensitivity of the supernova dynamics to the 
employed flux-limiting scheme must be expected (Messer~et~al.~\cite{mm98},
Lichtenstadt et al.~\cite{li98}).

Modifications of flux-limited diffusion have been suggested 
(Janka~\cite{ja91a}, \cite{ja92}; Dgani \& Janka~\cite{dg92}, 
Cernohorsky \& Bludman~\cite{cb94}) by which
considerable improvement can be achieved for spherically symmetric,
static and time-independent backgrounds (Smit~et~al.~\cite{sc97}), but 
satisfactory performance for the general time-dependent and
non-stationary case has not been demonstrated yet. Therefore the
interest turns towards direct solutions of the Boltzmann equation
for neutrino transport, also because the need to check the
applicability of any approximation with more elaborate methods
remains. Moreover, the rapid increase of the computer power and the wish to
become independent of ad hoc constraints on generality or accuracy
yield a motivation for the efforts of several groups (in particular
Mezzacappa \& Bruenn~\cite{me93a},\cite{me93b},\cite{me93c} and
Messer~et~al.~\cite{mm98}; more recently also Burrows~\cite{bu97})
to employ such Boltzmann solvers in 
neutrino-hydrodynamics calculations of supernova explosions.

There are different possibilities to solve the Boltzmann equation
numerically, one of which is by straightforward discretization of 
spatial, angular, energy, and time variables and conversion of the
differential equation into a finite difference equation which can 
then be solved for the values of the neutrino phase space distribution 
function at the discrete mesh points. Dependent on the number $N$ of 
angular mesh points, this procedure is called S$_{N}$ method. 
Since solving the equation is computationally very expensive,
there are limitations to the resolution in angle and energy space.
Therefore tests need to be done whether a chosen (and affordable)
number of energy and angle grid points is sufficient to describe
the spectra well and, in particular, to reproduce the highly 
anisotropic neutrino distribution outside the neutrinosphere.

Another, completely different approach to solve the Boltzmann equation
is the Monte Carlo (MC) method by which the probabilistic history
of a large number of sample neutrinos is followed to simulate the
neutrino transport statistically (Tubbs~\cite{tu78}; 
Janka~\cite{ja87}, \cite{ja91a};
Janka \& Hillebrandt~\cite{jh89a}). In principle, the accuracy of the 
results is only limited by the statistical fluctuations associated
with the finite number of sample particles. Since the MC transport
essentially does not require the use of angle and energy grids,
it allows one to cope with highly anisotropic angular distributions
and to treat with high accuracy neutrino reactions with an arbitrary 
degree of energy exchange between neutrinos and matter. However, the
MC method is also computationally very time consuming, in particular
if high accuracy on a fine spatial grid or at high optical depths 
is needed. Therefore it is
not the transport scheme of one's choice for coupling it with a 
hydrodynamics code.

In the present work, we make use of the advantages of the MC method 
in order to test the accuracy and reliability of a newly developed
neutrino transport code that follows the lines of the S$_N$ scheme 
described by 
Mezzacappa \& Bruenn~(\cite{me93a},\cite{me93b},\cite{me93c}). In particular,
we shall test the influence of the number of radial, energy, and angular 
mesh points on predicting the spectra and the radial evolution of the
neutrino flux in ``realistic'' protoneutron star atmospheres as found
in hydrodynamic simulations of supernova explosions (Wilson~\cite{wi88}).
Since our investigations are restricted to static and 
time-independent backgrounds, we concentrate on generic properties
of the transport description which should also hold for more general
situations. The radial evolution of the angular distribution of the
neutrinos is such a property, because it is primarily dependent on 
the profile of the opacity and the geometry of the neutrino-decoupling
region, but is not very sensitive to the details of the temperature
and composition in the neutron star atmosphere. Finally, good overall
agreement of the MC and S$_N$ results would strengthen the credibility
of the MC transport with its limited ability to yield high spatial 
resolution.

The paper is organized as follows. The details of the Boltzmann solver 
and essential information for the MC method are given in Sect.~2. 
In Sect.~3 we describe the background models. Section~4 presents the results
of our comparative calculations, i.e., neutrino spectra, luminosities,
and Eddington factors. Some of our calculations are also compared against 
results obtained with a MGFLD code developed by Suzuki~(\cite{su94}). The
dependence of the results from the S$_N$ scheme on the energy, angular,
and radial grid resolution is discussed, too. It is shown that an angular
mesh that varies with the position in the star can improve the angular
resolution and the representation of the beamed neutrino distributions 
without increasing the number of angular mesh points. Finally, a summary
of our results and a discussion of their implications can be found in 
Sect.~5.

\section{Numerical methods}

\subsection{Boltzmann solver}

\subsubsection{Basic equations}

Our S$_N$ code is based on a finite difference form of 
the general relativistic 
Boltzmann equation for neutrinos. We assume spherical symmetry of the
star throughout this paper. For the Misner-Sharp metric 
(Misner \& Sharp~\cite{msh64}):
\begin{eqnarray}
\label{eqn:mtrc}
{\rm d}s^{2}  = {\rm e} ^{2 \phi(t, m)}{\rm c}^{2} {\rm d}t^{2} 
& - &{\rm e} ^{2 \lambda(t, m)} 
\left(\frac{\rm G}{{\rm c}^{2}}\right)^{2} {\rm d}m^{2} 
\nonumber \\ & - &r^{2}(t, m) 
({\rm d}\theta^{2} + \sin^{2}\theta {\rm d}\phi^{2}) \quad ,
\end{eqnarray}
the Boltzmann equation in the Lagrangian frame takes the following form:
\begin{eqnarray}
\label{eqn:be}
\frac{1}{\rm c} {\rm e}^{- \phi} \frac{\partial}{\partial \, t} 
\left( \frac{f_{\nu}}{\rho_{\rm b}} \right)
& + & 4 \pi {\rm e}^{- \phi} \mu \frac{\partial \ {\rm e}^{\phi} r^{2} f_{\nu}}
{\partial \, m} \nonumber \\
& + & \frac{\partial}{\partial \mu} \left\{ 
(1 - \mu ^{2}) \left[ 2 \pi \rho_{\rm b} \frac{\partial r^{2}}{\partial m} 
\right. \right.
\nonumber \\
& + & \mu \frac{1}{\rm c} {\rm e}^{- \phi} 
\frac{\partial \ln (\rho_{\rm b} r^{3})}{\partial \, t} 
\nonumber \\
& - & \left. \left. 
4 \pi r^{2} \rho_{\rm b} \frac{\partial \phi}{\partial m} \right] 
\left( \frac{f_{\nu}}{\rho_{\rm b}} \right)\right\} \nonumber \\
& + & \left[ \mu ^{2} \frac{1}{\rm c} {\rm e}^{- \phi} 
\frac{\partial \ln (\rho_{\rm b} r^{3})}{\partial \, t}
\right .
\nonumber \\
& - & \frac{1}{\rm c} {\rm e}^{- \phi} \frac{\partial \ln r}{\partial \, t} 
\nonumber \\
& - & \left. \mu 4 \pi r^{2} \rho_{\rm b} 
\frac{\partial \phi}{\partial m} \right]
\frac{\partial \ \ }{\partial \frac{1}{3} \varepsilon _{\nu} ^{3}} 
\left\{ \varepsilon _{\nu} ^{3} \left( {\frac{f_{\nu}}{\rho_{\rm b}}} \right)
\right\} \nonumber \\
& = & \frac{1}{\rho_{\rm b}} \frac{1}{\rm c} {\rm e}^{- \phi} 
\left( \frac{\delta f_{\nu}}{\delta \, t} \right)_{\rm coll} \quad . 
\end{eqnarray}
In the above formula ${\rm c}$ and ${\rm G}$ are the velocity of light and the
gravitational constant, respectively, $t$ is the coordinate time and 
$m$ is the baryonic mass coordinate which is related to the
circumference radius $r$ through the conservation law of the baryonic 
mass. In view of combination with a Lagrangian hydrodynamics code
(Yamada~\cite{ya97}), the baryonic mass is chosen to be the independent 
variable instead of the radius. $\phi$, $\lambda$ and $r$ are 
the metric components which are determined by the Einstein equations. 
In this paper, however, these quantities are
given from the background models and set to be constant with time during the
neutrino transport calculations. $f_{\nu}$ is the neutrino phase space 
distribution function. Under the assumption of
spherical symmetry, $f_{\nu}$ is a function of $t$, $m$, $\mu$  and
$\varepsilon _{\nu}$, where $\mu$ is the cosine of the angle of the  
neutrino momentum with respect to the outgoing radial direction and
$\varepsilon _{\nu}$ is the neutrino energy. $\rho_{\rm b}$ is the
baryonic mass density. 
\par
The right hand side of 
Eq.~(\ref{eqn:be}) is the so-called collision term, which actually includes
absorption, emission, scattering and pair creation and annihilation of 
neutrinos, details of which are described below. The left
hand side, on the other hand, looks a little bit different from the
form used, for example, in 
Mezzacappa~et~al.~(\cite{me93a},\cite{me93b},\cite{me93c}). This is 
not only because it is fully general
relativistic but also because all the velocity dependent terms are
expressed as time derivatives so that it can easily be coupled
to the implicit general relativistic Lagrangian hydrodynamics code 
(Yamada~\cite{ya97}). In this way a fully coupled, implicit system of
the radiation-hydrodynamics equations is formed, 
in which the time derivatives can be 
treated easily because they are just off-diagonal components of the
matrix set up from the linearized equations. It should be noted, however, 
that since we assume 
that the matter background is static in this paper, all these 
time derivatives are automatically set to be zero, 
although these terms have been already implemented in the code. 
\par
The conserved neutrino number $N _{\nu}$ in the absence 
of source terms is represented in terms of the chosen independent
variables as 
\begin{equation}
\label{eqn:totnum}
N_{\nu} = \int f_{\nu} (t,\, m,\, \mu,\, \varepsilon_{\nu})
\ \frac{{\rm d}m}{\rho _{\rm b}} \,
\frac{2 \pi \varepsilon _{\nu} ^{2} {\rm d} \varepsilon _{\nu} \, {\rm d} \mu}
{({\rm hc}) ^{3}} \quad .
\end{equation}
This suggests to cast Eq.~(\ref{eqn:be}) in a conservation
form with respect to the neutrino number. It is also evident that the 
combination of $\left ( \displaystyle{\frac{f _{\nu}}{\rho _{\rm b}}}
\right ) $ is more
convenient to be used than $f _{\nu}$ itself. In the following, therefore, 
we define the specific neutrino distribution function 
as in Mezzacappa~et~al.~(\cite{me93a},\cite{me93b},\cite{me93c}) by  
\begin{equation}
\label{eqn:deff}
F _{\nu} \equiv \left ( \frac{f _{\nu}}{\rho _{\rm b}} \right )
\end{equation}
and use this quantity as the dependent variable to be solved for.

\subsubsection{Finite difference scheme of Boltzmann equations}

As mentioned above the specific neutrino distribution function $F _{\nu}$ is a
function of $t$, $m$, $\mu$ and $\varepsilon
_{\nu}$. This four-dimensional phase space is discretized and
Eq.~(\ref{eqn:be}) is written as a finite-difference equation. In the time 
direction we adopt a fully implicit differencing. The
discretized specific distribution function $F^{n}_{i, j, k}$ 
is defined at the mesh
centers of the spatial, angular and energy grids. Here the subscripts $i, j, 
k$ refer to the spatial, angular and energy grid points,
respectively. The superscript $n$ corresponds to the time step. The
value at each cell interface is evaluated by interpolation of  
the distribution at two adjacent mesh centers. 
\par 
Our finite difference method is essentially the same as that of 
Mezzacappa~et~al.~(\cite{me93a},\cite{me93b},\cite{me93c}) 
with some modifications. 
For the spatial advection, the upwind difference and the centered
difference are linearly averaged with the weights determined by the
ratio of the mean free path to the distance to the stellar surface 
unlike Mezzacappa~et~al.~(\cite{me93a},\cite{me93b},\cite{me93c}) 
who used the ratio of the mean free path to the local mesh width.   
In fact, in the latter case we found that the upwind distribution was
given too large a weight in 
the optically thick region and thus the flux was overestimated. 
This issue will be addressed later.
\par
The angular mesh is determined such that each mesh center and cell width 
correspond to the abscissas and weights of the Gauss-Legendre
quadrature, respectively. In angular direction, 
the neutrino distribution at each interface is simply taken
as the upwind value. 
The advection in the energy space is also approximated by an upwind
scheme following Mezzacappa \& Matzner~(\cite{me89}), although this
does not allow to conserve both lepton number and energy in non-static
situations (which are not considered here) unless a large number of 
energy zones is used (Mezzacappa, private communication).
\par
In typical calculations, 105, 6 and 12 mesh points are used for the spatial,
angular and energy discretizations, respectively. The dependence of 
the results on the numbers of mesh points will be discussed below. 
The finite-differenced Boltzmann equation forms a nonlinear coupled
system of equations for all radial grid points which  
is linearized and solved iteratively by using a 
Newton-Raphson scheme. The linearized equations adopt a block
tridiagonal matrix form, which can be efficiently solved by the Feautrier
method. 

\subsection{Monte Carlo method}

Different from the finite difference method, the Monte Carlo method
constructs the statistical ensemble average by following the destinies 
of individual test particles and performing the average when all
particles have been transported. Due to the fact that neutrinos are
fermions it is impossible to propagate them independently. Instead, 
the full time-dependent problem has to be simulated by following a large number
(typically $\sim500,000$) of sample particles along their trajectories 
simultaneously in order to be able to construct the local phase space
occupation functions and to include anisotropies as well as 
phase space blocking 
effects self-consistently into the calculation of the reaction rates
and source terms. The modeling of the phase space distribution 
function from the local particle sample must guarantee the correct
approach to chemical equilibrium. Also the Pauli
exclusion principle has to be satisfied by the statistical average. 
We refer readers to Janka~(\cite{ja87}) and references therein for details. 
\par
The stellar background is divided into 15 equispaced spherical shells
of homogeneous composition and uniform thermodynamical conditions, the 
number of which was determined both from physical requirements for
spatial resolution and from the requirement to have acceptably small  
statistical errors in the local neutrino phase space distributions
constructed from the chosen number of sample particles. 
Although the Monte Carlo method 
is essentially mesh free, about 60 energy bins and approximately 35 angular
bins are used only for representing the phase space distribution functions and 
for calculating the reaction kernels. Neutrinos are injected into the
computational volume at the inner boundary in the way described in 
the next section, while particles passing inwardly through 
the inner boundary  
are simply forgotten. The outer boundary is treated as a free boundary, where
particles escape unhindered and no neutrino is assumed to come in from 
outside.
 
\subsection{Boundary conditions}
\label{sec_boundary}

In the presented calculations we are mainly interested in the neutrino 
transport in the region where the neutrinos decouple from the stellar
background and the emitted spectra form. Therefore we calculate the
neutrino transport only in the vicinity of the ``neutrinosphere''. For 
this reason we have to set an inner
boundary condition as well as an outer boundary condition for each
model. At the outer boundary we impose the condition 
that no neutrinos enter the computational volume from outside. In the
Boltzmann solver, this is realized by setting 
\begin{equation}
f_{\nu}(r_{\rm s}, \ \mu, \ \varepsilon _{\nu}) = \left \{
\begin{array}{ll}
f_{\nu}(r_{\rm max}, \ \mu, \ \varepsilon _{\nu}) & 
\quad  {\rm for} \ \mu \ge 0
\\ 0 &
\quad {\rm for} \ \mu < 0   , 
\end{array} \right .
\end{equation} 
where $r_{\rm max}$ is the radius of the outermost mesh center, and
$r_{\rm s}$ is the radius of the outer surface which is dislocated outward
from the outermost mesh center by half a radial cell width.
\par
On the other hand, we have to specify the
distribution of the neutrinos coming into the computational volume 
at the inner boundary $r_{\rm IB}$ which is
dislocated inward from the innermost mesh center $r_{\rm min}$ by 
half a radial cell width. 
For this purpose we adopt  
Fermi-Dirac distribution functions, in which the temperature, chemical
potential and a normalization factor are determined such that 
the neutrino number density at $r_{\rm IB}$ (where the neutrinos are
essentially isotropically distributed because the inner boundary is
chosen to be located at high optical depth), 
the average energy $\left\langle \varepsilon _{\nu}
\right\rangle$ and the width of 
the energy spectrum, measured by the parameter $\displaystyle{
\frac{\left\langle \varepsilon _{\nu} ^{2} \right\rangle}
{\left\langle \varepsilon _{\nu} \right\rangle^2}}$ 
(see Janka \& Hillebrandt~1989\cite{jh89b}),  are reproduced as given 
by Wilson's (\cite{wi88}) models. Thus the inner boundary 
condition is set as :
\begin{equation}
\label{eqn:ibc}
f_{\nu}(r_{\rm IB}, \ \mu, \ \varepsilon _{\nu}) = \left \{
\begin{array}{l}
\displaystyle{\frac{1}{A}} \frac{1}{{\rm e} ^{(\varepsilon _{\nu} -
\mu _{\rm IB}) / ({\rm k_{B}} T_{\rm IB})} + 1} \\ 
\quad \quad \quad \quad \quad \quad \quad \quad {\rm for} \ \mu \ge 0 \\
f_{\nu}(r_{\rm min}, \ \mu, \ \varepsilon _{\nu}) \\
\quad \quad \quad \quad \quad \quad \quad \quad {\rm for} \ \mu < 0
\quad , 
\end{array} \right .
\end{equation}
where $A$, $\mu _{\rm IB}$ 
and $T_{\rm IB}$ are the fitting parameters, 
the values of which are summarized 
in Table~\ref{tab2} for all considered models and neutrino species. Concerning
the distribution of neutrinos that leave the computational volume, we impose in
Eq.~(\ref{eqn:ibc}) in the Boltzmann solver 
the condition that it is the same as the 
distribution at the innermost mesh center. From the physical point of
view, however, and treated correctly in the Monte Carlo simulations, 
it should be determined by the fraction of neutrinos which is emitted 
or backscattered towards the inner boundary. This is in
general different from what the phase space distribution at the 
innermost mesh center yields because the mean free path of 
the neutrinos near the inner boundary 
is shorter than the mesh width. As a result the imposed condition 
for $\mu < 0$ in Eq.~(\ref{eqn:ibc}) leads to a minor discrepancy of
the treatment of the inner boundary condition in the Monte Carlo and
Boltzmann computations and sometimes causes a small 
oscillation of the neutrino distribution near the inner boundary in
the latter computations. 
This issue will be revisited later. 
\begin{table}
\caption[ ]{Fitting parameters for the Fermi-Dirac distributions 
which describe the spectra at the inner boundary.\label{tab2}}
\begin{flushleft}
\begin{tabular}{ccrrr}
\noalign{\smallskip}
\hline 
\noalign{\smallskip} 
models & neutrinos & $T_{\rm IB}$ [MeV] & $\mu _{\rm IB}$ [MeV] & 
\multicolumn{1}{c}{$A$} \\ 
\noalign{\smallskip} 
\hline 
\noalign{\smallskip} 
W1 & $\nu _{e}$       &  9.56932  &  7.85190 &  1.221490 \\
   & $\bar{\nu} _{e}$ &  9.38259  &  7.13610 &  3.874140 \\
   & $\nu _{\mu}$     & 10.32539  & 30.87195 & 17.911456 \\
W2 & $\bar{\nu} _{e}$ & 10.52454  &  5.75932 &  3.672680 \\
W3 & $\bar{\nu} _{e}$ &  9.87875  & 11.27370 &  7.108076 \\
\noalign{\smallskip} 
\hline 
\end{tabular}
\end{flushleft}
\end{table}

\subsection{Input physics}

\subsubsection{Neutrino reactions}

The neutrino opacities of dense neutron star matter are still one of
the major uncertainties of supernova simulations. Theoretical and
numerical complications arise from the description and treatment of 
nucleon thermal motion and recoil (Schinder~\cite{sc90}), nuclear force 
effects and nucleon blocking (Prakash~et~al.~\cite{pl97}; 
Reddy~et~al.~\cite{rp97}),
and nucleon correlations, spatial (Sawyer~\cite{sa89}; 
Burrows \& Sawyer~\cite{bs98a},\cite{bs98b}; Reddy~et~al.~\cite{rp98a}) 
as well as temporal (Raffelt \& Seckel~\cite{ra95}; 
Raffelt~et~al.~\cite{ra96}).
Although in particular nucleon recoil and auto-correlations might
play an important role even in the sub-nuclear outer layers of the 
protoneutron star down to densities below $10^{13}\,{\rm g\,cm}^{-3}$
(Janka~et~al.~\cite{jk96}; Hannestad \& Raffelt~\cite{hr97}) 
we do not concentrate
on this problem here but rather employ the standard description of
the neutrino opacities, according to which neutrinos interact with 
isolated nucleons (see, e.g., Tubbs \& Schramm~\cite{ts75}; 
Bruenn~\cite{br85}).
Also, as in most other simulations, bremsstrahlung production of
neutrino-antineutrino pairs is neglected here, although it may be
important as pointed out by Suzuki~(\cite{su93}) and more recently by
Burrows~(\cite{bu97}) and Hannestad \& Raffelt~(\cite{hr97}). 
Doing so, we intend to enable comparison with other
(already published) work and want to avoid the mixing of effects from
a different numerical treatment of the transport with those from a
non-standard description of neutrino-nucleon interactions or from the
inclusion of processes typically not considered in the past.
\par
The following neutrino reactions have been implemented in our codes.    
\\[1ex]
[1] $\nu _{e} + n  \rightleftharpoons e + p$ \\
\hspace*{0.3cm}
(electron-type neutrino absorption on neutrons), \\[1ex]
[2] $\bar{\nu} _{e} + p  \rightleftharpoons e^{+} + n$ \\
\hspace*{0.3cm}
(electron-type anti-neutrino absorption on protons), \\[1ex]
[3] $\nu + N  \rightleftharpoons \nu + N$ \\
\hspace*{0.3cm}
(neutrino scattering on nucleons), \\[1ex]
[4] $\nu + e \rightleftharpoons \nu + e$ \\
\hspace*{0.3cm}
(neutrino scattering on electrons). \\[1ex]
In addition to the above reactions, the following processes have also been 
implemented in both codes, although these reactions are not used in
the present paper.
\\[1ex]
[5] $\nu _{e} + A \rightleftharpoons A + e^{-}$ \\
\hspace*{0.3cm}
(electron-type neutrino absorption on nuclei), 
\\[1ex]
[6] $\nu + A \rightleftharpoons \nu + A $ \\
\hspace*{0.3cm}
(neutrino coherent scattering on nuclei), 
\\[1ex]
[7] $e^{-} + e^{+} \rightleftharpoons \nu + \bar{\nu} $ \\
\hspace*{0.3cm}
(electron-positron pair annihilation and creation), 
\\[1ex]
[8] $\gamma ^{\ast} \rightleftharpoons \nu + \bar{\nu} $ \\
\hspace*{0.3cm}
(plasmon decay and creation). \\[1ex]
Since the pair processes are not taken into account in this paper, we
can treat each species of neutrinos separately. A test showed that in
the considered protoneutron star atmospheres neutrino pair creation
and annihilation as well as processes involving nuclei 
are unimportant to determine the fluxes and spectra.  

\subsubsection{EOS}

In this paper we use a simplified equation of state, in which only
nucleons, electrons, alpha particles and photons are included. They
are all treated as ideal gases. For given density, temperature and
electron fraction we derive the mass fractions and the chemical 
potentials of nucleons and the electron chemical potential from this
equation of state. The disregard of nuclei is well justified 
for the densities and
temperatures we are considering, where most of the nuclei are
dissociated into free nucleons. This is actually confirmed by comparing 
our EOS with the more realistic EOS of Wolff that is based on 
the Skyrme-Hartree-Fock method (Hillebrandt~\&~Wolff~\cite{hw85}). 
Only very small differences of the nucleon 
chemical potentials are found for the innermost region where 
small amounts of nuclei appear and for the outermost region 
where some contribution from alpha particles is mixed into the stellar 
medium. We also 
repeated some of the calculations making use of the Wolff EOS 
with the nuclei-related
reactions [5], [6] (neutrino emission, absorption and scattering on 
nuclei) turned on and found qualitatively and quantitatively 
the same results. Hence the nuclei-related
reactions are switched off in the calculations described below.

\section{Models}

\subsection{Stellar models}

The time-dependent transport calculations presented here were performed 
for background profiles which are representative of protoneutron
star atmospheres during the quasi-static neutrino cooling phase
(Wilson~\cite{wi88}). At this stage, several seconds after core bounce, the 
typical evolution timescale of density, temperature, and electron fraction
is much longer than the timescale for neutrinos to reach a stationary
state. Therefore our assumption of a static and time-independent
background is justified. In addition, our interest is focused on
the radial evolution of the Eddington factors and on a test of
the influence of the energy and angle resolution used in the S$_N$ 
Boltzmann solver. Both aims do not require a fully self-consistent 
approach which takes into account the evolution of the stellar
background (in particular of the temperature and composition). In fact,
the Eddington factors are normalized angular integrals of the radiation
intensity and as such reflect very general characteristics of 
the geometrical structure of the atmosphere where neutrinos and matter
decouple. 

\begin{figure}

% for preprint

\epsfxsize=8.8cm  \epsffile{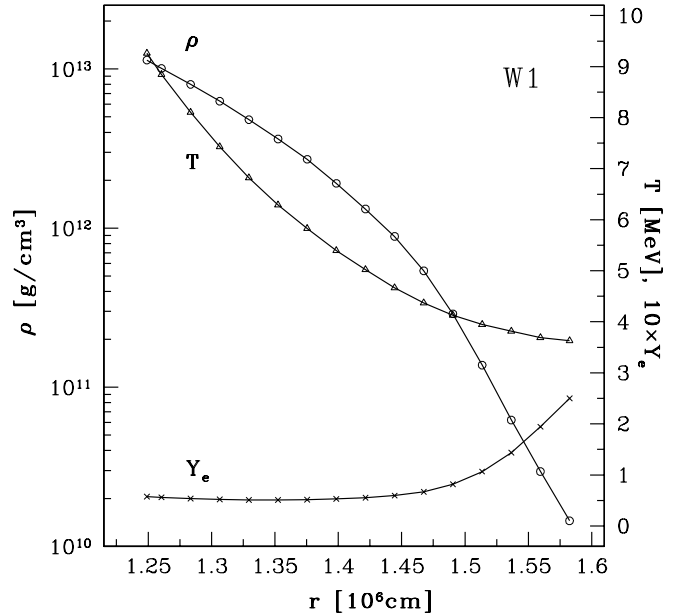} 
\caption[]{The profile of density, temperature and electron
fraction for Wilson's (1988) post bounce core model W1 ($t =
3.32\,$s).}
\label{fig_1}

% for submission

%\picplace{1cm}
%\caption{The profile of density, temperature and electron
%fraction for Wilson's (1988) post bounce core model W1 ($t =
%3.32\,$s). 
%\label{fig_1}}
\end{figure}

\begin{figure}
\epsfxsize=8.8cm  \epsffile{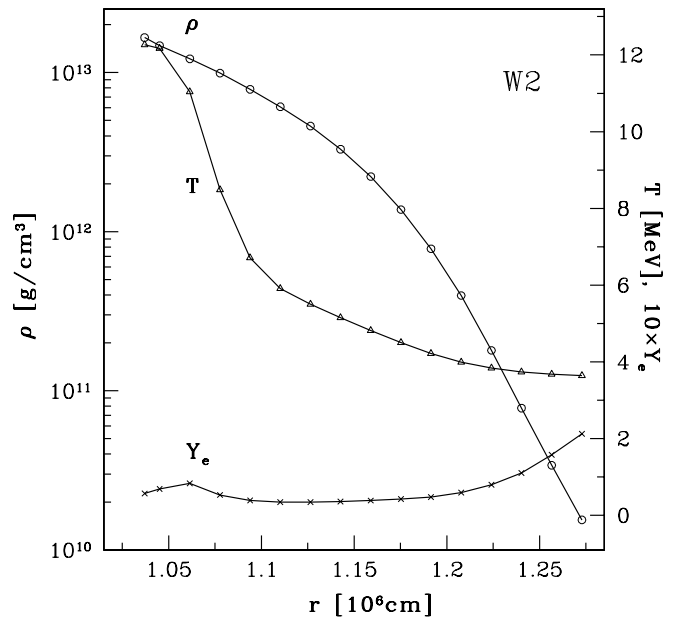} 
\caption[]{The profile of density, temperature and electron
fraction for Wilson's post bounce core model W2 ($t = 5.77\,$s).}
\label{fig_2}

%\picplace{1cm}
%\caption{The profile of density, temperature and electron
%fraction for Wilson's post bounce core model W2 ($t = 5.77\,$s). 
%\label{fig_2}}
\end{figure}

\begin{figure}
\epsfxsize=8.8cm  \epsffile{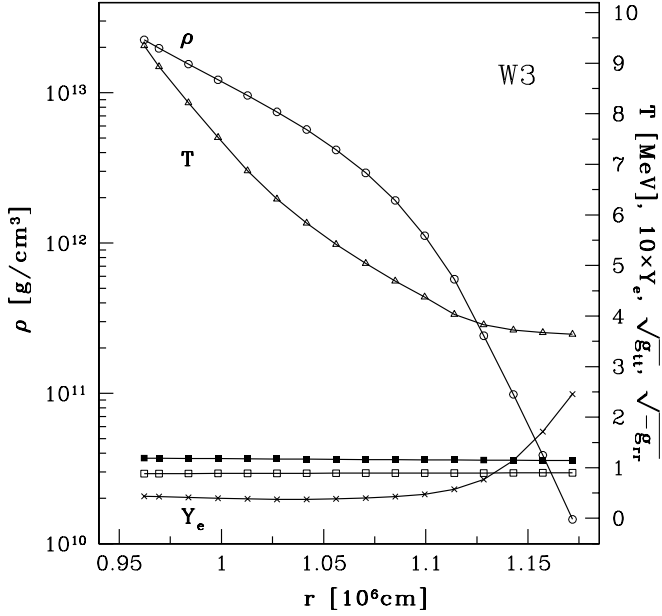} 
\caption[]{The profile of density, temperature and electron
fraction as well as the metric coefficients $\sqrt{g_{tt}}$ (open
squares) and $\sqrt{-g_{rr}}$ (filled squares) for Wilson's post 
bounce core model W3 ($t = 7.81\,$s).}
\label{fig_3}

%\picplace{1cm}
%\caption{The profile of density, temperature and electron
%fraction as well as the metric coefficients $\sqrt{g_{tt}}$ and 
%$\sqrt{-g_{rr}}$ for Wilson's post bounce core model W3 ($t =
%7.81\,$s). \label{fig_3}}
\end{figure}

Profiles from Wilson's (\cite{wi88}) protoneutron star model were taken for 
three different times, 3.32, 5.77, and 7.81~s after core bounce.
With the chosen fundamental variables density $\rho$, temperature 
$T$, and electron fraction $Y_e$, the thermodynamical state is defined 
for the plasma consisting of non-relativistic free nucleons, 
arbitrarily relativistic and degenerate electrons and positrons,
and photons in thermal equilibrium. Figures~\ref{fig_1}--\ref{fig_3} show
the input used for the three models. In Fig.~\ref{fig_3} also the 
general relativistic metric coefficients $\sqrt{g_{tt}} = e^{\phi}$ and
$\sqrt{-g_{rr}} = e^{\lambda}$ are given as provided by Wilson's data
and used for a comparative general relativistic calculation of the
neutrino transport in model~W3.

\subsection{Computed transport models}

All models computed with the Boltzmann code are summarized 
in Table~\ref{tab3}. Models ST are
the standard models, in which 105 uniform spatial, 6 angular and 12 
energy mesh points were used. The energy mesh is
logarithmically uniform and covers 0.9 -- 110 MeV.  
The numbers of angular grid points and energy grid points were increased in
models~FA and FE, respectively. In model~CS we used the same radial
grid as in the Monte Carlo simulations where 15 radial zones were
chosen. 105 spatial mesh points were again
used in model~NI with no interpolations of density, temperature and
electron fraction in the radial grid of the Monte Carlo
simulations. Model~GR took into account the general relativistic
effects. We used a non-uniform spatial mesh in model~NU. A
different interpolation of up-wind differencing and centered
differencing was tried for the radial advection term in model~DI. 
We assumed the nucleon scattering to be isotropic in model~IS. 
As is understood from Table~\ref{tab3}, most of the comparisons were
done for the electron-type anti-neutrinos, since they are most
important from the observational point of view.     

\begin{table*}
\caption[ ]{Characteristics of the calculated models. 
$105_{r} \times 6_{\mu} \times 12_{\varepsilon}$ implies that 105
spatial mesh points, 6 angular mesh points and 12 energy mesh points 
are used. See the text for details. \label{tab3}}
\begin{flushleft}
\begin{tabular}{clccl}
\noalign{\smallskip}
\hline
\noalign{\smallskip}
model & \multicolumn{1}{c}{mesh} & background model & $\nu$ species & 
\multicolumn{1}{c}{notes} \\
\noalign{\smallskip}
\hline
\noalign{\smallskip}
ST1 & $105_{r} \times 6_{\mu} \times 12_{\varepsilon}$  & W1 & 
$\nu_{e}$       & interpolation in spatial grid of MC
\\
ST2 & $105_{r} \times 6_{\mu} \times 12_{\varepsilon}$  & W1 & 
$\bar{\nu} _{e}$ & interpolation in spatial grid of MC
\\
ST3 & $105_{r} \times 6_{\mu} \times 12_{\varepsilon}$  & W2 & 
$\bar{\nu} _{e}$ & interpolation in spatial grid of MC
\\
ST4 & $105_{r} \times 6_{\mu} \times 12_{\varepsilon}$  & W3 & 
$\bar{\nu} _{e}$ & interpolation in spatial grid of MC
\\
ST5 & $105_{r} \times 6_{\mu} \times 12_{\varepsilon}$  & W1 & 
$\nu _{\mu}$     & interpolation in spatial grid of MC
\\
FA  & $105_{r} \times 10_{\mu} \times 12_{\varepsilon}$ & W1 & 
$\bar{\nu} _{e}$ & interpolation in spatial grid of MC
\\
FE  & $105_{r} \times 6_{\mu} \times 18_{\varepsilon}$  & W1 & 
$\bar{\nu} _{e}$ & interpolation in spatial grid of MC
\\
CS  & $15_{r} \times 6_{\mu} \times 12_{\varepsilon}$   & W1 & 
$\bar{\nu} _{e}$ & same spatial grid as in MC
\\
NI  & $105_{r} \times 6_{\mu} \times 12_{\varepsilon}$  & W1 & 
$\bar{\nu} _{e}$ & 
no interpolation in spatial grid of MC  \\
GR  & $105_{r} \times 6_{\mu} \times 12_{\varepsilon}$  & W3 & 
$\bar{\nu} _{e}$ & 
general relativity included \\
NU  & $105_{r} \times 6_{\mu} \times 12_{\varepsilon}$  & W1 & 
$\bar{\nu} _{e}$ & 
non-uniform spatial mesh  \\
DI  & $105_{r} \times 6_{\mu} \times 12_{\varepsilon}$  & W1 & 
$\nu_{\mu}$ & 
different interpolation for conservative radial advection\\
IS  & $105_{r} \times 6_{\mu} \times 12_{\varepsilon}$  & W1 & 
$\bar{\nu} _{e}$ & 
isotropic $\nu$-$N$ scattering \\
\noalign{\smallskip}
\hline
\end{tabular}
\end{flushleft}
\end{table*}

\section{Numerical results}

\subsection{Luminosity and average energy \label{la}}

In the following, we consider the neutrino transport results after the
time-dependent simulations have reached steady states.

First we compare observable quantities such as the
luminosity, average energy and average squared energy of the neutrino 
flux for models ST and GR.  These quantities are calculated at the 
outermost spatial zone as
\begin{eqnarray}
\label{eqn:lum}
L_{\nu} & = & 4 \pi r_{\rm s}^2 {\rm c} \int   
\frac{2 \pi \varepsilon _{\nu} ^{2} {\rm d} \varepsilon _{\nu} {\rm d} \mu}
{({\rm hc}) ^{3}} \ 
f_{\nu} \,  \mu \, \varepsilon _{\nu} \quad \quad ,\\
\label{eqn:emin}
\left\langle \varepsilon _{\nu} \right\rangle & = & 
\displaystyle{\frac{\displaystyle 
{\int \frac{2 \pi \varepsilon _{\nu} ^{2} 
{\rm d} \varepsilon _{\nu} {\rm d} \mu}
{({\rm hc}) ^{3}}} \ 
f_{\nu} \,  \mu \,  \varepsilon _{\nu}}{\displaystyle
{\int \frac{2 \pi \varepsilon _{\nu} ^{2} 
{\rm d} \varepsilon _{\nu} {\rm d} \mu}
{({\rm hc}) ^{3}}} \ 
f_{\nu} \,  \mu }} \quad \quad ,\\
\label{eqn:e2min}
\left\langle \varepsilon _{\nu} ^{2} \right\rangle & = & \displaystyle{\frac{
\displaystyle{\int \frac{2 \pi \varepsilon 
_{\nu} ^{2} {\rm d} \varepsilon _{\nu} {\rm d} \mu}
{({\rm hc}) ^{3}}} \ 
f_{\nu} \,  \mu \,  \varepsilon _{\nu} ^{2}}{\displaystyle{\int   
\frac{2 \pi \varepsilon _{\nu} ^{2} {\rm d} \varepsilon _{\nu} {\rm d} \mu}
{({\rm hc}) ^{3}}} \ 
f_{\nu} \,  \mu }} \quad \quad ,
\end{eqnarray} 
where $r_{\rm s}$ is again the surface radius, 
and the redshift corrections from the surface up to infinity are not
taken into account. The results are summarized in Table~\ref{tab4}. 

\begin{table*}
\caption[ ]{Luminosity, average energy and average squared energy
for models~ST1--ST5 and for model~GR. B and MC in the second column 
refer to the Boltzmann simulations and the Monte Carlo simulations, 
respectively. For the definitions of $L _{\nu}$, 
$\left\langle \varepsilon _{\nu} \right\rangle$ and 
$\left\langle \varepsilon _{\nu} ^{2} \right\rangle$ see 
Eqs.~(\ref{eqn:lum}), (\ref{eqn:emin}) and (\ref{eqn:e2min}) in the text.
\label{tab4}}
\begin{flushleft}
\begin{tabular}{cccccccc}
\noalign{\smallskip}
\hline
\noalign{\smallskip}
 &  &
$\begin{array} {c} \nu _{e} \\ \rm{W1} \end{array}$ & 
$\begin{array} {c} \bar{\nu}_{e} \\ \rm{W1} \end{array}$ & 
$\begin{array} {c} \bar{\nu}_{e} \\ \rm{W2} \end{array}$ &
$\begin{array} {c} \bar{\nu}_{e} \\ \rm{W3} \end{array}$ & 
$\begin{array} {c} \nu _{\mu} \\ \rm{W1} \end{array}$ &
$\begin{array} {c}  \bar{\nu}_{e} \\ \rm{GR, W3} \end{array}$\\
\noalign{\smallskip}
\hline
\noalign{\smallskip}
$L _{\nu}$                & B  & $7.1 \times 10^{51}$ & 
$4.0 \times 10^{51}$ & $2.5 \times 10^{51}$ & 
$1.5 \times 10^{51}$ & $4.9 \times 10^{51}$ & $1.3 \times 10^{51}$ \\
\footnotesize{[erg/sec]}  & MC & $7.3 \times 10^{51}$ &
$4.0 \times 10^{51}$ & $2.6 \times 10^{51}$ & 
$1.5 \times 10^{51}$ & $4.9 \times 10^{51}$ & $1.3 \times 10^{51}$ \\
$\left\langle \varepsilon _{\nu} \right\rangle$ 
& B  & 12.8 & 16.3 & 15.9 & 16.2 & 24.3 & 15.5 \\
\scriptsize{[MeV]}     & MC & 12.7 & 16.1 & 15.8 & 16.3 & 24.2 & 15.6 \\
$\left\langle \varepsilon _{\nu} ^{2} \right\rangle$ 
& B  & 198.5 & 329.0 & 309.9 & 324.5 &
727.8 & 300.3 \\
\scriptsize{[MeV$^{2}$]}    & MC & 198.6 & 322.6 & 308.9 & 327.1 &
724.3 & 300.6 \\
\noalign{\smallskip}
\hline
\end{tabular}
\end{flushleft}
\end{table*}

The electron-type neutrino has the lowest energy while the muon-type
neutrino has the highest. The reason for this is that the electron-type  
neutrino has the shortest mean free path due to absorptions on
the abundant neutrons, and decouples in the surface-near layers where
the temperature is lower. In contrast, the muon and tau neutrinos do
not interact with particles of the stellar medium by charged currents
and therefore their thermal decoupling occurs at a higher
temperature. The luminosity for the
electron-type anti-neutrino gets smaller as time passes, which is
due to the cooling and shrinking of the protoneutron star. 
As can be seen in the table, the agreement of all quantities 
between the two methods is very good, which confirms the statistical
convergence of the Monte Carlo simulations. The average energy is in general
determined accurately because the energy spectrum is 
shaped in the region where the neutrino angular distribution is
not very anisotropic. Moreover, possible effects due to the rather
coarse  angular resolution of the Boltzmann code essentially  
cancel out by taking the ratios of Eqs.~(\ref{eqn:emin}) and 
(\ref{eqn:e2min}). On the other hand, the luminosity and 
the number flux of the Monte Carlo computations are also well
reproduced by the Boltzmann results. This is due to the fact 
that these quantities are also determined deep inside the star and 
are nearly conserved farther out. In fact, when integrating  
Eq.~(\ref{eqn:be}) over angle and energy multiplying with unity and 
$\varepsilon _{\nu}$, respectively, ignoring all time derivatives and 
general relativistic effects, one gets 
\begin{eqnarray}
\label{eqn:nexch}
\frac{\partial F^{n}_{\nu} (r)}{\partial \ m \quad} & = & 
\frac{1}{\rho _{\rm b}} \int \frac{2 \pi \varepsilon _{\nu} ^{2} 
{\rm d} \varepsilon _{\nu} {\rm d} \mu}{({\rm hc}) ^{3}} \ 
\left( \frac{\delta f_{\nu}}{\delta \, t} \right)_{\rm coll} \quad , \\
\label{eqn:eexch}
\frac{\partial L^{\varepsilon}_{\nu} (r)}{\partial \ m \quad} & = & 
\frac{1}{\rho _{\rm b}} \int \frac{2 \pi \varepsilon _{\nu} ^{2} 
{\rm d} \varepsilon _{\nu} {\rm d} \mu}
{({\rm hc}) ^{3}} \ \varepsilon _{\nu} \,
\left( \frac{\delta f_{\nu}}{\delta \, t} \right)_{\rm coll} \quad ,
\end{eqnarray}  
where $F^{n}_{\nu} (r)$ and $L^{\varepsilon}_{\nu} (r)$ are the number 
and energy fluxes of neutrinos at radius $r$, respectively, and
defined as,
\begin{eqnarray}
F_{\nu}^{n} (r) & = & 4 \pi r^2 {\rm c} \int   
\frac{2 \pi \varepsilon _{\nu} ^{2} {\rm d} \varepsilon _{\nu} {\rm d} \mu}
{({\rm hc}) ^{3}} \ 
f_{\nu} \,  \mu \quad , \\
L_{\nu}^{\varepsilon} (r) & = & 4 \pi r^2 {\rm c} \int   
\frac{2 \pi \varepsilon _{\nu} ^{2} {\rm d} \varepsilon _{\nu} {\rm d} \mu}
{({\rm hc}) ^{3}} \ 
f_{\nu} \,  \mu \, \varepsilon _{\nu} 
\quad .
\end{eqnarray}
As expected intuitively, the scattering kernels drop out of 
the right hand side of Eq.~(\ref{eqn:nexch}) 
while only the isoenergetic scattering does not contribute 
on the right hand side of Eq.~(\ref{eqn:eexch}). 
In the given Boltzmann code Eqs.~(\ref{eqn:nexch}) and
(\ref{eqn:eexch}) are discretized in a conservative form for the
radial advection. Therefore it is clear that the Boltzmann code can
calculate the number and energy fluxes accurately if also the number
and energy exchange by the reactions are calculated accurately in 
the source terms on the right hand sides of the equations. 
Moreover, the radial evolutions of $F_{\nu}^{n} (r)$ and 
$L_{\nu}^{\varepsilon} (r)$ are entirely determined by the number
and energy exchange through reactions of which the net effect is 
small in an atmospheric layer
which is in a stationary state and reemits as much energy and lepton
number as it absorbs. Changes of the number fluxes and luminosities in 
the considered protoneutron star atmospheres occur only in regions
where the neutrino distribution is still essentially isotropic and
possible effects due to an insufficient angular resolution in the
Boltzmann S$_N$ scheme do not cause problems. 
For all these reasons it is not surprising that the same quality 
of agreement is also found for the radial evolutions 
of the luminosity, average energy and average squared energy and that
this is also true for the general relativistic case. 
 
\subsection{Neutrino energy spectra}

In the previous section we discussed only the energy and angle
integrated quantities. However, we also provide information about the energy
spectra of each neutrino species, because they yield more evidence
about the quality of the agreement between the calculations with the
different codes. 
\par
\begin{figure}
\epsfxsize=8.8cm  \epsffile{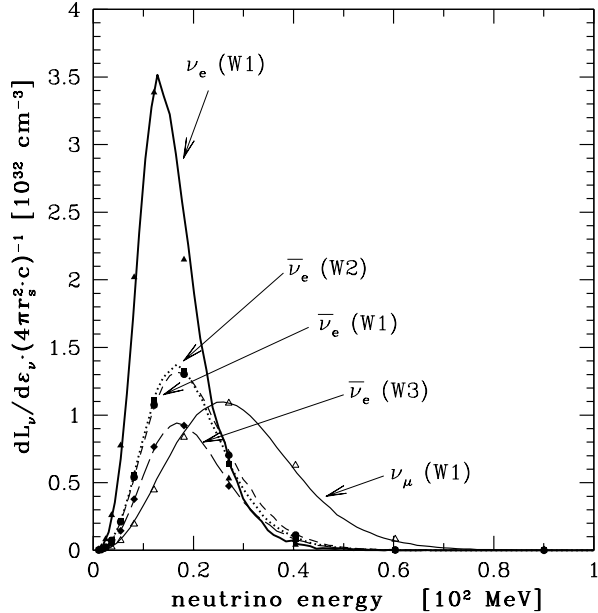} 
\caption[]{The neutrino energy flux spectra at the protoneutron star
surface. The lines show the results 
of the Monte Carlo simulations (the thick solid line,
the short dashed line and the thin solid line for electron-type
neutrinos, electron-type anti-neutrinos and muon-type neutrinos,
respectively, in case of background model W1, and the dotted line
and the long dashed line for electron-type anti-neutrinos in case of
background models W2 and W3, respectively). The symbols represent the
results of the corresponding Boltzmann simulations (the filled 
triangles, the filled circles and the open triangles for electron-type
neutrinos, electron-type anti-neutrinos and muon-type neutrinos,
respectively, in case of background model W1, and the filled 
squares and the filled diamonds for electron-type anti-neutrinos in case of
background models W2 and W3, respectively).}
\label{fig_4}

%\picplace{1cm}
%\figurenum{4}
%\caption{The neutrino energy flux spectra at the protoneutron star
%surface. The lines show the results 
%of the Monte Carlo simulations (the thick solid line,
%the short dashed line and the thin solid line for electron-type
%neutrinos, electron-type anti-neutrinos and muon-type neutrinos,
%respectively, in case of background model W1, and the dotted line
%and the long dashed line for electron-type anti-neutrinos in case of
%background models W2 and W3, respectively). The symbols represent the
%results of the corresponding Boltzmann simulations (the filled 
%triangles, the filled circles and the open triangles for electron-type
%neutrinos, electron-type anti-neutrinos and muon-type neutrinos,
%respectively, in case of background model W1, and the filled 
%squares and the filled diamonds for electron-type anti-neutrinos in case of
%background models W2 and W3, respectively).  \label{fig_4}}
\end{figure}
\begin{figure}
\epsfxsize=8.8cm  \epsffile{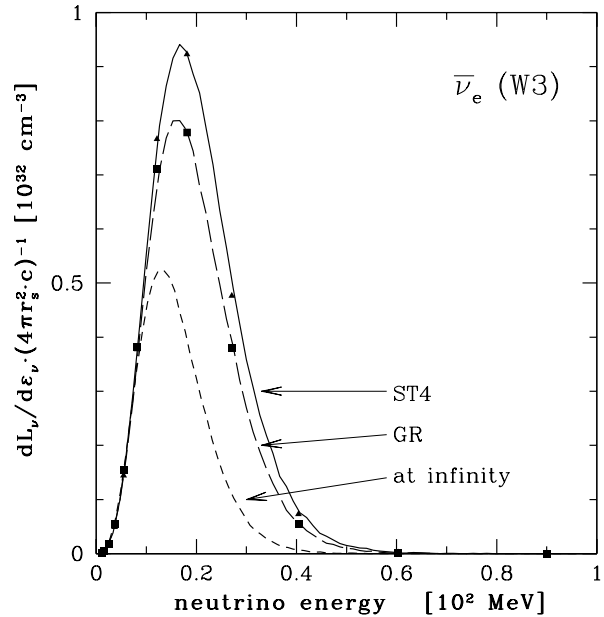} 
\caption[]{The electron-type anti-neutrino energy flux spectra for background
model W3 with and without general relativity. The long 
dashed line and the filled squares show the results of
the Monte Carlo simulation and the Boltzmann simulation, respectively,
with the general relativistic effects included. The solid line and the filled
triangles are for the corresponding non-relativistic results. 
The short dashed line is the redshift-corrected energy spectrum at
infinity calculated from the Monte Carlo results, the other results
are given at the protoneutron star surface.}
\label{fig_5}

%\picplace{1cm}
%\figurenum{5}
%\plotfiddle{fig_5.ps}{16cm}{0}{80}{80}{-250}{-80}
%\caption{The electron-type anti-neutrino energy flux spectra for background
%model W3 with and without general relativity. The long 
%dashed line and the filled squares show the results of
%the Monte Carlo simulation and the Boltzmann simulation, respectively,
%with the general relativistic effects included. The solid line and the filled
%triangles are for the corresponding non-relativistic results. 
%The short dashed line is the redshift-corrected energy spectrum at
%infinity calculated from the Monte Carlo results, the other results
%are given at the protoneutron star surface.
%\label{fig_5}}
\end{figure}

Figs.~\ref{fig_4} and \ref{fig_5} show energy flux spectra defined as
\begin{equation}
\frac{1}{4 \pi r_{s}^{2} \, {\rm c}}\frac{dL_{\nu}}{d\varepsilon_{\nu}}
 = \int \frac{2 \pi {\rm d} \mu}{({\rm hc}) ^{3}} \ 
f_{\nu} \,  \mu \, \varepsilon _{\nu}^{3} 
\quad ,
\end{equation}
at the protoneutron star surface for different cases. For the
reasons discussed in section~\ref{la}, the spectra computed with the
Boltzmann code (symbols) and the Monte Carlo code (lines) show excellent 
agreement. The
number and the distribution of the energy bins in the Boltzmann code
seem to be adequate to reproduce the highly resolved Monte Carlo
spectra.   

\subsection{Flux factor and Eddington factor \label{fefac}} 

So far we discussed only angle integrated quantities since they are
observable. However we are also interested in the angular
distributions of the neutrinos, because information about the angular
distributions is important to determine the neutrino heating rate 
in the hot-bubble region (see Eq.~(\ref{eq-1})). 
Although it is an advantage of the Boltzmann solver 
over MGFLD that one does not have to assume an ad hoc closure relation
between the angular moments of the distribution function,
one should remember that the usable number of angular mesh
points is severely limited. In the Feautrier method 
the computation time increases in proportion to the third power of 
the dimension of the blocks in the tridiagonal block matrix 
which has to be inverted when one chooses the radius as the 
outermost variable of the do-loops. The dimension of one block, 
in turn, is linearly proportional to the number of angular mesh
points. The same dependence holds for the number of 
energy mesh points and the number of neutrino species. In the 
standard calculations we use 6 angular
mesh points and 12 energy grid points, and we treat a single neutrino 
species at a time, which corresponds to a block matrix size of 72. On the
other hand, the number of spatial grid points is about 100. 
The CPU time is a few 
seconds per inversion of the whole matrix on a single vector processor 
of a Fujitsu VPP500. Hence, use of more than 10 angular mesh points is almost
prohibitive for calculations with three neutrino species even with a 
highly parallelized matrix inversion routine
(Sumiyoshi~\&~Ebisuzaki~\cite{se98}). It is, therefore,
important to clarify the sensitivity of the accuracy to the  
angular resolution. 
\par 
For this reason we consider the flux factor 
$\left\langle \mu \right\rangle _{\rm e1}$ and the Eddington 
factor $\left\langle \mu ^{2} \right\rangle _{\rm e1}$ which are defined as~:
\begin{eqnarray}
\label{eqn:ffac}
\left\langle \mu \right\rangle _{\rm e1} & = & 
\displaystyle{\frac{\displaystyle 
{\int \frac{2 \pi \varepsilon _{\nu} ^{2} 
{\rm d} \varepsilon _{\nu} {\rm d} \mu}
{({\rm hc}) ^{3}}} \ 
f_{\nu} \,  \mu \,  \varepsilon _{\nu}}{\displaystyle
{\int \frac{2 \pi \varepsilon _{\nu} ^{2} 
{\rm d} \varepsilon _{\nu} {\rm d} \mu}
{({\rm hc}) ^{3}}} \ 
f_{\nu} \,  \varepsilon _{\nu}}} \quad \quad , \\
\left\langle \mu ^{2} \right\rangle  _{\rm e1} & = & 
\displaystyle{\frac{\displaystyle 
{\int \frac{2 \pi \varepsilon _{\nu} ^{2} 
{\rm d} \varepsilon _{\nu} {\rm d} \mu}
{({\rm hc}) ^{3}}} \ 
f_{\nu} \,  \mu ^{2} \,  \varepsilon _{\nu}}{\displaystyle
{\int \frac{2 \pi \varepsilon _{\nu} ^{2} 
{\rm d} \varepsilon _{\nu} {\rm d} \mu}
{({\rm hc}) ^{3}}} \ 
f_{\nu} \,  \varepsilon _{\nu}}} \quad \quad .
\end{eqnarray} 
Here the subscript ``$\rm e1$'' means that the averages are defined with
the weight of energy. In MGFLD these factors 
are related with each other by a closure
condition which can be derived from the employed flux-limiter
(Janka~\cite{ja91a}, \cite{ja92}). For simplicity we discuss here
only energy integrated quantities as defined above. The fundamental
features are similar for the individual energy groups. 
\par
\begin{figure}
\epsfxsize=8.8cm  \epsffile{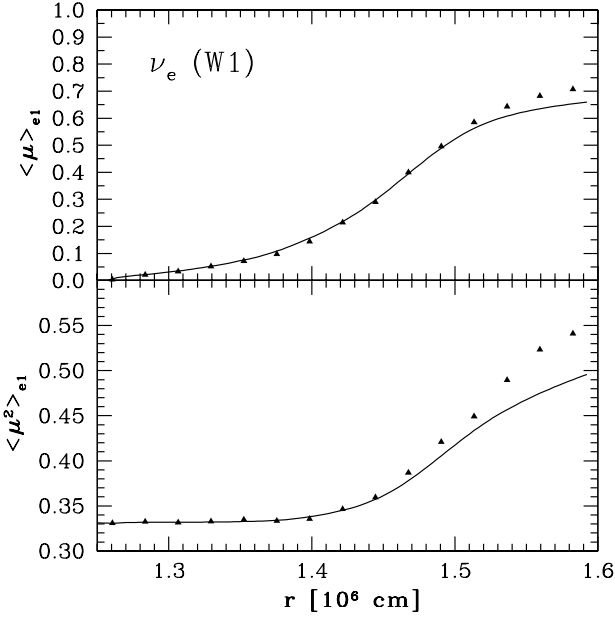} 
\caption[]{The flux factor (upper panel) and the Eddington factor
(lower panel) of the electron-type neutrino for model W1. 
The filled triangles and the solid line show the results of
the Monte Carlo simulation and the Boltzmann simulation, respectively.}
\label{fig_6}

%\picplace{1cm}
%\figurenum{6}
%\caption{The flux factor (upper panel) and the Eddington factor
%(lower panel) of the electron-type neutrino for model W1. 
%The filled triangles and the solid line show the results of
%the Monte Carlo simulation and the Boltzmann simulation, respectively.
%\label{fig_6}}
\end{figure}
\begin{figure}
\epsfxsize=8.8cm  \epsffile{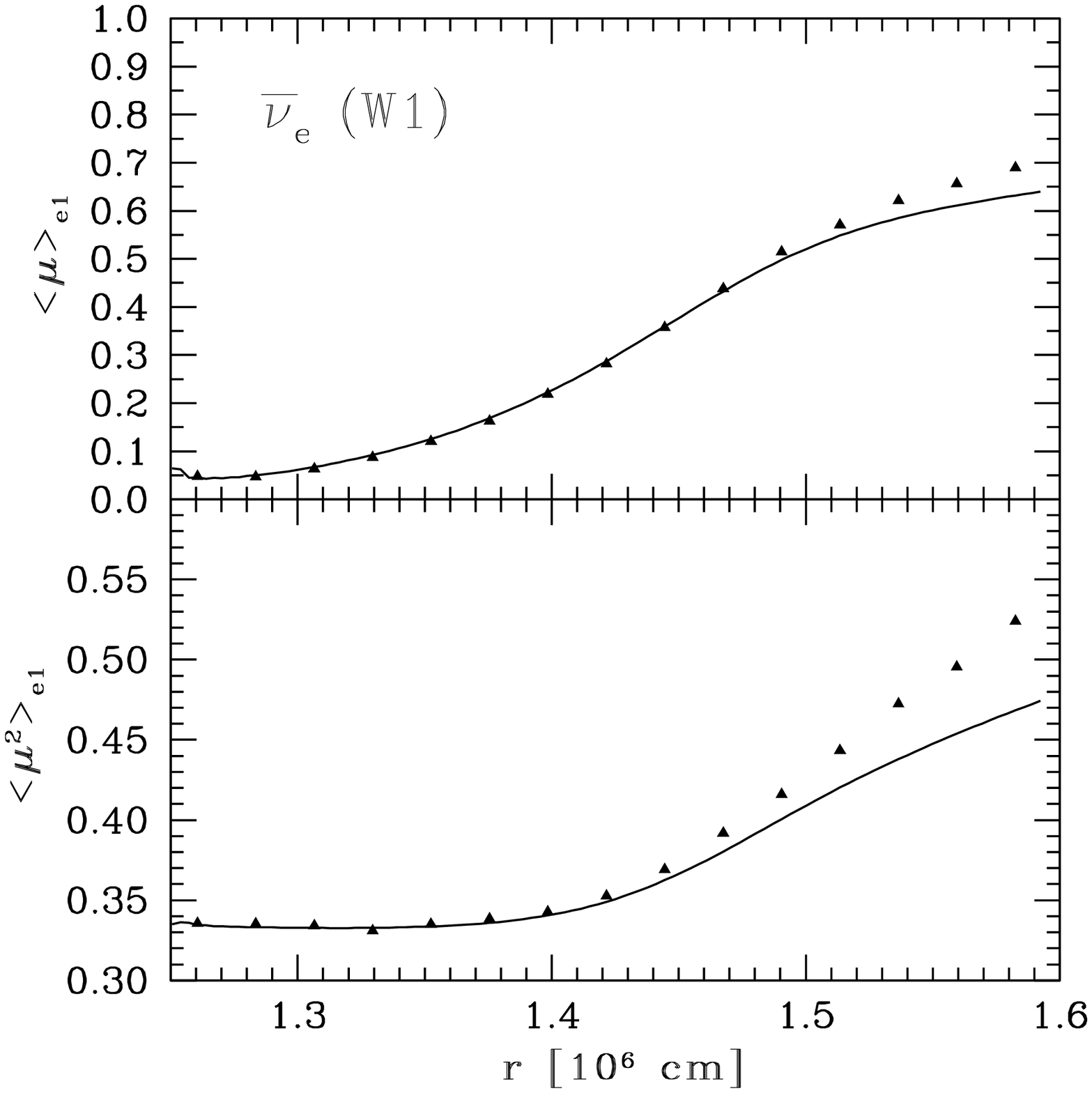} 
\caption[]{The same as Fig.~\ref{fig_6} but for the electron-type 
anti-neutrino.}
\label{fig_7}

%\picplace{1cm}
%\figurenum{7}
%\caption{The same as Fig.~\ref{fig_6} but for the electron-type 
%anti-neutrino. 
%\label{fig_7}}
\end{figure}
\begin{figure}
\epsfxsize=8.8cm  \epsffile{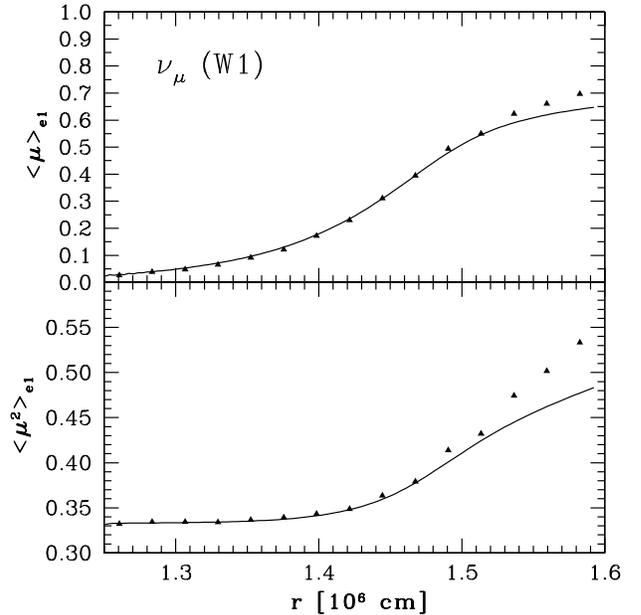} 
\caption[]{The same as Fig.~\ref{fig_6} but for the muon-type 
neutrino.}
\label{fig_8}

%\picplace{1cm}
%\figurenum{8}
%\caption{The same as Fig.~\ref{fig_6} but for the muon-type 
%neutrino.
%\label{fig_8}}
\end{figure}

In Figs.~\ref{fig_6}--\ref{fig_8}, 
we show the radial evolutions of the flux factors and the 
Eddington factors for all neutrino species in case of background model~W1. 
The upper panels show the flux factors and the lower panels 
the corresponding Eddington factors. The solid lines are the results 
of the Boltzmann simulations (having the finer radial resolution) 
and the filled triangles are those  
of the Monte Carlo simulations. As can be seen, near the inner 
boundary the flux factors are almost zero while the Eddington factors 
are $1/3$, which implies that the neutrino angular distribution is 
nearly isotropic, a consequence of the fact that the neutrinos are 
in equilibrium with the surrounding 
matter. As we move outward, both factors begin to deviate from these 
values, reflecting the increase of the mean free path and a more rapid
diffusion. Farther out, the angular moments increase monotonically
towards unity, the value in the free 
streaming limit, as the angular distribution gets more and more forward
peaked with increasing distance from the source. As can be seen clearly in
Figs.~\ref{fig_6}--\ref{fig_8}, the Boltzmann solver tends to 
{\it underestimate} both
angular moments in the outermost region, where the neutrino angular 
distribution is most forward peaked. This can be explained by 
the insufficient angular resolution, or, to be more specific, by the
fact that in case of the employed Gauss-Legendre quadrature the maximum 
angle cosine $\mu_{\rm max}$ of the angular grid is significantly less than
unity, if not a large number of angular grid points is used. This is
directly confirmed by using a larger number of angular mesh points
or, in particular, a variable angular mesh (see Sect.~\ref{vam}).
\par
\begin{figure}
\epsfxsize=8.8cm  \epsffile{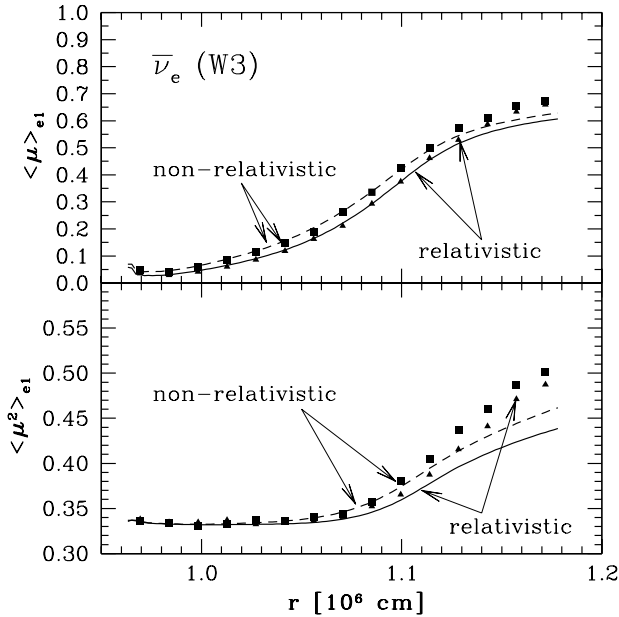} 
\caption[]{The flux factor (upper panel) and the Eddington factor 
(lower panel) of the electron-type anti-neutrino for model~GR 
where the general relativistic effects are taken into account. The
solid line represents the results of the Boltzmann simulation while
the solid triangles correspond to those of the Monte Carlo
simulation. For comparison the corresponding non-relativistic results,
model~ST4, are shown with the solid squares and the dashed lines for 
the Monte Carlo simulation and the Boltzmann simulation, respectively.}
\label{fig_9}

%\picplace{1cm}
%\figurenum{9}
%\caption{The flux factor (upper panel) and the Eddington factor 
%(lower panel) of the electron-type anti-neutrino for model~GR 
%where the general relativistic effects are taken into account. The
%solid line represents the results of the Boltzmann simulation while
%the solid triangles correspond to those of the Monte Carlo
%simulation. For comparison the corresponding non-relativistic results,
%model~ST4, are shown with the solid squares and the dashed lines for 
%the Monte Carlo simulation and the Boltzmann simulation, respectively. 
%\label{fig_9}}
\end{figure}

The same trend is also present in the general relativistic case, model~GR,
as shown in Fig.~\ref{fig_9}. It turned out that the ray bending effect
which tends to isotropize the angular distribution of the neutrinos is not 
very important and that differences between the Monte Carlo results and
the Boltzmann results are significantly larger.
\par
\begin{figure}
\epsfxsize=8.8cm  \epsffile{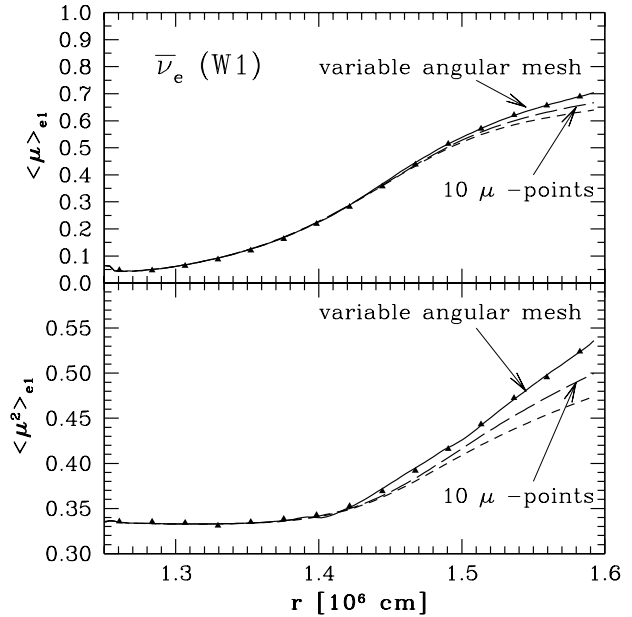} 
\caption[]{The flux factors (upper panel) and the Eddington factor 
(lower panel) of the electron-type anti-neutrino for background
model~W1 with the long dashed lines for model~FA where 10 angular mesh points 
are used instead of 6. The short dashed lines show for comparison 
the results of the corresponding standard model~ST2. The solid lines 
are the result obtained with the variable angular mesh method. 
The Monte Carlo result is shown with the solid triangles.}
\label{fig_10}

%\picplace{1cm}
%\figurenum{10}
%\caption{The flux factors (upper panel) and the Eddington factor 
%(lower panel) of the electron-type anti-neutrino for background
%model~W1 with the long dashed lines for model~FA where 10 angular mesh points 
%are used instead of 6. The short dashed lines show for comparison 
%the results of the corresponding standard model~ST2. The solid lines 
%are the result obtained with the variable angular mesh method. 
%The Monte Carlo result is shown with the solid triangles.
%\label{fig_10}}
\end{figure}

\begin{figure}
\epsfxsize=8.8cm  \epsffile{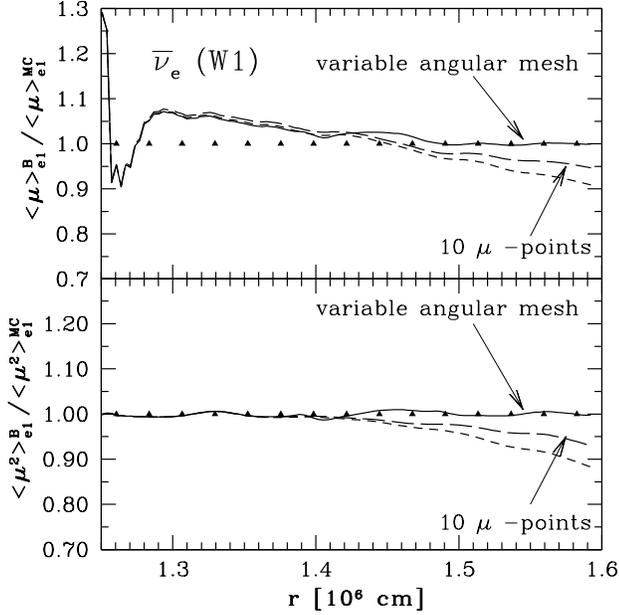}
\caption[]{Same as Fig.~\ref{fig_10} but showing the ratios of the
Boltzmann to the Monte Carlo results for the flux factor (upper panel)
and the Eddington factor (lower panel). Again, the Boltzmann calculations
with 6 angular bins (standard model~ST2, short dashed lines), 
10 angular bins (model~FA, long dashed lines) and the variable angular mesh
(solid lines) are shown. Between the radial grid points of the Monte
Carlo simulation (whose locations are indicated by solid triangles), 
the Monte Carlo results are interpolated by cubic splines.}
\label{fig_10m}
\end{figure}

 
In Fig.~\ref{fig_10} we show the flux factor and the Eddington factor 
for model~FA where we employed 10 angular mesh points instead of 6. 
The long dashed lines are for model~FA and the short dashed lines
depict the result of model~ST2, the corresponding standard model, for 
comparison. The discrepancy between the Boltzmann simulation and 
the Monte Carlo simulation is reduced with the increase of the number 
of angular mesh points. This supports our interpretation that
the deviation stems entirely from the insufficient angular resolution 
and/or the unfavorable location of the angular grid points for the
Gauss-Legendre quadrature used in
the Boltzmann simulations. Indeed, the degree of improvement is
consistent with the fact that our finite difference scheme is of first
order for the angular advection, since we always take the upwind 
differencing as explained above. Even with the finer 10-zone angular mesh 
the deviation of both factors from the exact values given by the Monte 
Carlo result is significant. The maximum deviations,
$1-\mu_{\rm max}$ for the flux factor and $1-\mu_{\rm max}^2$ for the
Eddington factor, are approached as one goes farther out into the 
optically thin regime. This is visible in Fig.~\ref{fig_10m} which displays
the ratios of the Boltzmann to the Monte Carlo results for the flux factors
and the Eddington factors in case of 6 and 10 angular bins and the variable 
angular mesh. (The relatively large discrepancies of the flux factors
for smaller radii are explained by slight differences of the treatment
of the inner boundary condition, see Sect.~\ref{sec_boundary}, and by the 
fact that the flux factor adopts very small values in the optically
thick region.) It should be mentioned that the flux factors
calculated in the recent paper by Messer~et~al.~(\cite{mm98}) 
do not converge to unity but saturate at a nearly constant lower 
level (around 0.9) even far outside of the neutrinosphere. This
reflects the use of $\mu_{\rm max} = 0.93$ for the largest $\mu$-bin of
the angular grid in the 6-point quadrature of the S$_6$ method.
\par
\begin{figure}
\epsfxsize=8.8cm  \epsffile{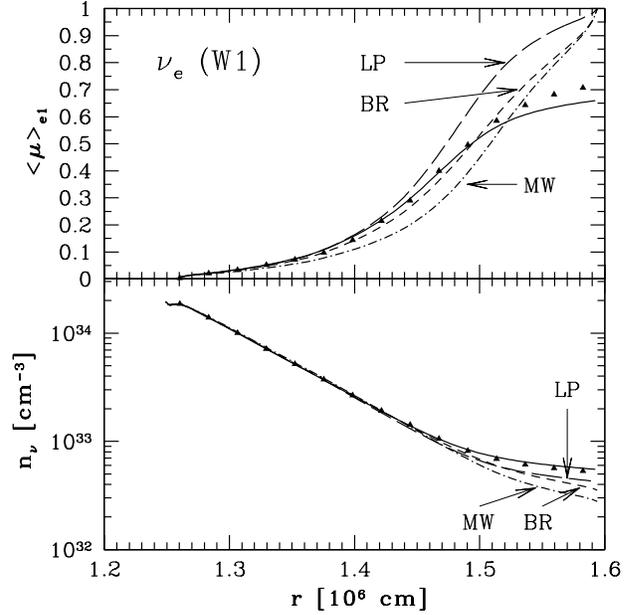} 
\caption[]{The flux factors and number densities of the electron-type 
neutrino as obtained by MGFLD with three different flux limiters,
Bruenn's (BR) with the short dashed lines, Levermore \& Pomraning's (LP)
with the long dashed lines, and Mayle \& Wilson's (MW) with the dash-dotted
lines. The background model is W1. For comparison, the Monte Carlo
result and the Boltzmann result (model ST1) are also plotted with the 
triangles and the solid lines, respectively.}
\label{fig_11}

%\picplace{1cm}
%\figurenum{11}
%\caption{The flux factors and number densities of the electron-type 
%neutrino of MGFLD with three different flux limiters,
%Bruenn's (BR) with the short dashed lines, Levermore \& Pomraning's (LP)
%with the long dashed lines, and Mayle \& Wilson's (MW) with the dash-dotted
%lines. The background model is W1. For comparison, the Monte Carlo
%result and the Boltzmann result are also plotted with the triangles and 
%the solid lines, respectively.
%\label{fig_11}}
\end{figure}
\begin{figure}
\epsfxsize=8.8cm  \epsffile{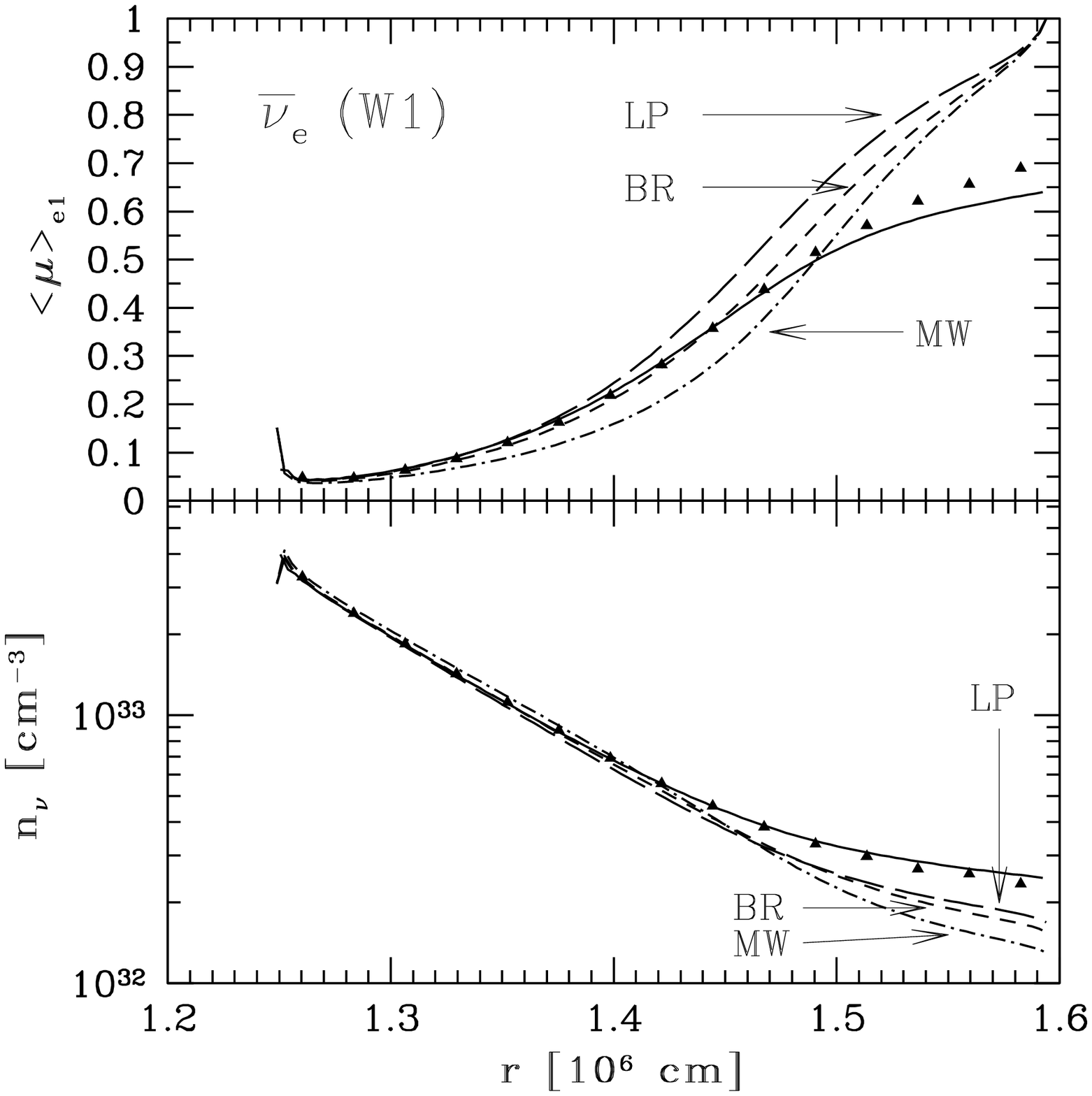} 
\caption[]{The same as Fig.~\ref{fig_11} but for the electron-type
anti-neutrino.} 
\label{fig_12}

%\picplace{1cm}
%\figurenum{12}
%\caption{The same as Fig.~\ref{fig_11} but for the 
%electron-type anti-neutrino.
%\label{fig_12}}
\end{figure}

It is interesting to see that this tendency of 
the Boltzmann solver is completely opposite to that of 
MGFLD. Janka~(\cite{ja91a}, \cite{ja92}) pointed out that all 
flux limiters used so far tend to overestimate the flux factor and
the Eddington factor in the optically thin region, which implies 
that the neutrino angular distribution approaches the free streaming 
limit much too rapidly (see also Messer et al.~\cite{mm98}). 
%added
In order to confirm this statement, transport calculations with MGFLD
were done for the same models with three different flux
limiters, which are Bruenn's (BR),
Levermore \& Pomraning's (LP) and Mayle \& Wilson's (MW). We refer readers 
to Janka~(\cite{ja92}) and Suzuki~(\cite{su94}) and references 
therein for details on the flux limiters. 
We show in Figs.~\ref{fig_11} and \ref{fig_12} the flux 
factors and local number densities of the electron-type neutrino and
electron-type anti-neutrino for model W1, respectively. It is
clear that all flux limiters overestimate the forward peaking of 
the angular distributions of the neutrinos in the optically thin
region, a trend that holds for all neutrino species and is not
dependent on the background model. The
typical deviation of MGFLD results from the Monte Carlo results is 
much larger than that between the Boltzmann solver and the Monte Carlo method. 

\par
\begin{figure}
\epsfxsize=8.8cm  \epsffile{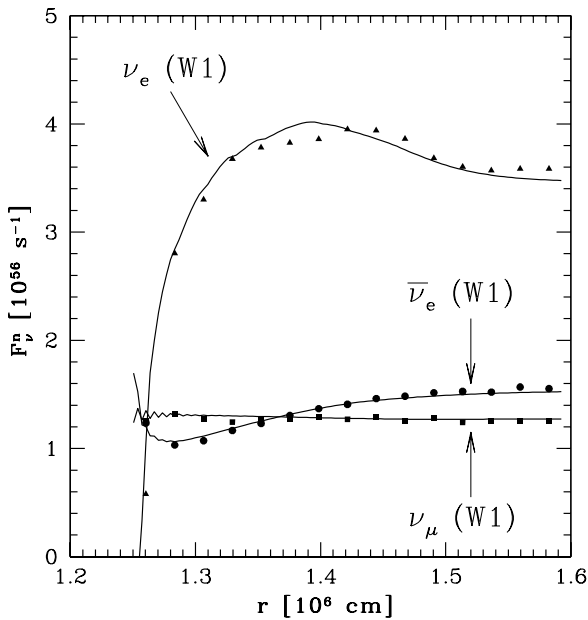} 
\caption[]{The number fluxes of all types of neutrinos for background 
model~W1. The filled symbols and the solid lines show the results of
the Monte Carlo simulations and the Boltzmann simulations, respectively.}
\label{fig_13}

%\picplace{1cm}
%\figurenum{13}
%\caption{The number fluxes of all types of neutrinos for model~W1. 
%The filled triangles and the solid lines show the results of
%the Monte Carlo simulations and the Boltzmann simulations, respectively.
%\label{fig_13}}
\end{figure}

From the lower panels of Figs.~\ref{fig_11} and \ref{fig_12} we learn 
that the local neutrino number density, which is given by
\begin{equation}
n_{\nu} (r)  =  \displaystyle 
{\int \frac{2 \pi \varepsilon _{\nu} ^{2} 
{\rm d} \varepsilon _{\nu} {\rm d} \mu}
{({\rm hc}) ^{3}}} \ f_{\nu} \quad , 
\end{equation}
is overestimated in case of the Boltzmann solver (by about 10\%) and 
underestimated for MGFLD (by approximately 30\%) in the optically 
thin region. This is understood from the fact that the number flux, 
which is defined as
\begin{equation}
\label{eqn:nflx}
F_{\nu}^{n}(r)  =  4 \pi r^{2} {\rm c} \displaystyle 
{\int \frac{2 \pi \varepsilon _{\nu} ^{2} 
{\rm d} \varepsilon _{\nu} {\rm d} \mu}
{({\rm hc}) ^{3}}} \ 
f_{\nu} \,  \mu    \quad ,
\end{equation} 
is related to the local neutrino number density by
\begin{equation}
\label{eqn:nfac}
n_{\nu} (r)  =  \displaystyle{
\frac{F_{\nu}^{n}(r)}{4 \pi r^{2} {\rm c} 
\left\langle \mu \right\rangle _{\rm e0}}
} \quad .
\end{equation}
$\left\langle \mu \right\rangle _{\rm e0}$ denotes the average angle 
cosine for the neutrino number flux and is calculated from Eq.~(\ref{eqn:ffac})
with a factor $\varepsilon_{\nu}$ omitted under the integrals in the
numerator and denominator.
Since the number flux is determined deep inside the protoneutron star 
atmosphere where the neutrinos are still nearly isotropic, and conserved
farther out, it is not affected by problems with a coarse angular
resolution occurring in the optically thin regime. 
For this reason, the number and energy fluxes agree well between 
the Monte Carlo method and the Boltzmann solver  
irrespective of the angular resolution as long as the Boltzmann solver is 
based on conservative finite differencing in the radial direction. 
The good agreement of the fluxes is confirmed by Fig.~\ref{fig_13}, 
which depicts the radial behavior of the number flux in case of 
models~ST1, ST2 and ST5. 
\par
It is now clear from Eq.~(\ref{eqn:nfac}) that an underestimation 
(overestimation) of the flux factor leads to an 
overestimation (underestimation ) of the number density, if
the flux is the same. It should be noted that even in the outermost
zone of our computational region in the atmosphere of a protoneutron
star, the neutrino angular distribution is not so strongly forward 
peaked as in the hot-bubble region farther out. Hence it must be
expected that the errors by an over- or underestimation of the neutrino number
density might be even larger in the hot-bubble region where neutrino
heating takes place. 
\par
Since the neutrino heating rate is proportional to the local neutrino
number density (actually: energy density)---this is why the
inverse of the flux factor appears in Eq.~(\ref{eq-1})---the
application of the Boltzmann solver with a relatively small number of
angular mesh points may lead to an overestimation of 
the neutrino energy deposition in the hot bubble for disadvantageous 
situations, in particular when most of the heating occurs in those 
regions where the deviation of the flux factor from the correct value
is significant. In contrast, all flux limiters used in MGFLD
underestimate the heating rate significantly. Therefore, even
for the Boltzmann solver, improvement of the angular resolution or a
redistribution of the angular grid points is desirable in order to
a priori avoid inaccurate evaluation of the heating rate. 
As already mentioned, an increase of the number of angular mesh points 
is not feasible. Choosing a variable angular mesh which adjusts mesh
point locations in dependence of time and spatial position might 
be a solution. 
%added
This issue will be discussed in the next subsection.

\subsection{Variable angular mesh in the Boltzmann solver \label{vam}}

Here we attempt to improve the angular resolution of 
the Boltzmann solver by redistributing the angular
grid points in dependence of time and position so that their density
is enhanced in the forward direction in the optically 
thin region where the neutrino angular distribution becomes strongly 
forward peaked and the Boltzmann solver underestimates the flux
factor and the Eddington factor. This requires adding extra 
angular advection terms in the numerical scheme 
which compensate for the motions of the angular mesh points. 
\par
We assume that the position of 
each interface of the angular grid is a function of time, baryonic 
mass and neutrino energy, i.e., $\mu_{\rm I} = \mu_{\rm I}(t,m,\varepsilon_{\nu})$. 
Integrating Eq.(\ref{eqn:be}) over angular bins then leads to the 
following additional angular advection fluxes at each angular mesh 
interface I~:
\begin{equation}
\label{eq:vat}
-\frac{1}{\rm c} \left( \frac{\partial \, \mu _{\rm I}}{\partial \, t} \right) 
\left (\frac{f_{\nu}}{\rho_{\rm b}} \right) \quad ,
\end{equation} 
\begin{equation}
\label{eq:vam}
-4 \pi \mu _{\rm I} \left(\frac{\partial \, \mu _{\rm I}}{\partial \, m} \right) 
\ {\rm e}^{\phi} \ r^{2} \  f_{\nu} \quad ,
\end{equation} 
\begin{eqnarray}
\label{eq:vae}
-\left[ \frac{1}{\rm c} \frac{\partial 
\left( \ln \rho_{\rm b} r^{3} \right)}{\partial \ t} \ 
\mu _{\rm I} ^{2} \right.
& - & {\rm e}^{\phi} \ 4 \pi r^{2} \ \rho_{\rm b} \frac{\partial \, 
\phi}{\partial \, m} \ \mu _{\rm I}  \nonumber \\
& - & \left. \frac{1}{\rm c} \frac{\partial \left( \ln r \right )}
{\partial \ t} \right]
\left( \frac{\partial \ \mu _{\rm I}}{\partial \frac{1}{3} \varepsilon _{\nu} ^{3}}
\right) \left (\frac{f_{\nu}}{\rho_{\rm b}} \right) .
\end{eqnarray} 
It is easy to understand that Eqs.~(\ref{eq:vat})--(\ref{eq:vae})
originate from the variability of the angular mesh points because of
the differentials of the $\mu _{\rm I}$ with respect to time, mass and
neutrino energy.  
\par
Since the neutrino reaction rates are strongly energy
dependent and so is the neutrino angular distribution, it would be desirable 
to implement the energy dependent angular mesh according to
Eq.~(\ref{eq:vae}). In the current preliminary attempt, however, we installed
only Eqs.~(\ref{eq:vat}) and (\ref{eq:vam}) for simplicity. 
Incidentally, since 
Eq.~(\ref{eq:vae}) is proportional to $\frac{\partial \phi}{\partial
m}$ in static background calculations, it is anyway negligible for the 
models considered here. We note that the motion of mesh points is 
not calculated implicitly, that is, the angular mesh points for 
the next time step are determined from the neutrino angular 
distribution at the current time step and are  
kept fixed during the implicit calculation of the transport for the next step. 
\par 
In Fig.~\ref{fig_10}, we show both the flux factor
and the Eddington factor obtained from the computation with the 
variable angular mesh. The improvement is clear from a comparison  
with the result of model~FA which employed 10 angular mesh
points and is also shown in the figure. 
It should be emphasized that increasing the number of angular mesh
points from 6 to 10 leads to an increase of CPU time by a factor of
$\sim$4.5, while the additional operations for the variable angular
mesh imply negligible computational load. 
\par
We repeated all Boltzmann calculations with the variable angular 
mesh method and found that the same improvement could be achieved for all
cases. Our scheme is stable at least for the static background models,
although the stability for dynamical background models remains to be
tested. Thus we think this method is promising in applying the
Boltzmann solver to the study of neutrino heating in the hot-bubble
region of supernovae, although there is room for
improvement concerning the prescription of the motion of the mesh points 
and the implementation of the energy-dependent angular mesh.

\subsection{Spatial and energy resolution in the Boltzmann solver 
and radial advection}

In this section we discuss how the numerical results change
in dependence of the number of spatial and energy grid points, 
the boundary condition and the treatment of the radial advection in
the Boltzmann solver. 
\par
\begin{figure}
\epsfxsize=8.8cm  \epsffile{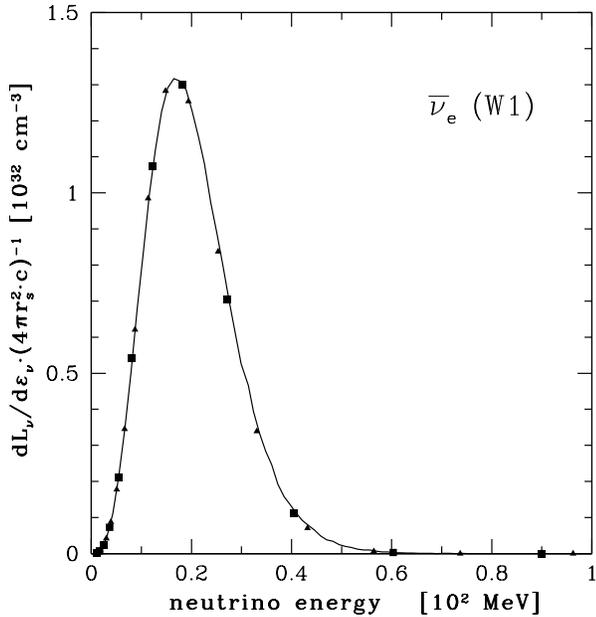} 
\caption[]{The energy spectrum of electron-type anti-neutrinos calculated
with 18 energy mesh points (model~FE, filled triangles) 
for background model W1. The solid line shows 
the result of the Monte Carlo simulation, and the solid squares
represent the result for the corresponding standard model~ST2 with
only 12 energy grid points.}
\label{fig_14}

%\picplace{1cm}
%\figurenum{14}
%\caption{The electron-type anti-neutrino energy spectrum calculated
%with 18 energy mesh points (model~FE, filled triangles) 
%instead of 12 for background model W1. The solid line shows 
%the result of the Monte Carlo simulation, and the solid squares
%represent the result for the corresponding standard model~ST2.
%\label{fig_14}}
\end{figure}

 
In Fig.~\ref{fig_14} we show the energy spectrum of electron-type
anti-neutrinos for
model~FE, in which we used 18 energy mesh points, compared with the
spectrum for the corresponding standard model ST2 which has 12 energy zones.
No qualitative or quantitative difference is found between the 
two cases. This is also true for the luminosity and
the angular distribution. Thus we think that about 15 energy mesh points
are sufficient for the calculation of the energy spectrum. These results
are in agreement with previous findings by 
Mezzacappa \& Bruenn~(\cite{me93a},\cite{me93b},\cite{me93c}).

\par
In models~CS and NI we reduced the spatial resolution, because 
the Monte Carlo simulations were done with only 15 radial mesh points 
which were used
to represent the stellar background on which the reaction kernels were 
evaluated. Another motivation for testing the sensitivity to the radial
resolution is that it is hardly possible to describe the protoneutron
star atmosphere with about 100 radial grid points in the context of a full
supernova simulation. 
In model~CS we used the same 15 spatial grid points as in the Monte
Carlo simulations. On the other hand, in model~NI 
we used 105 spatial mesh points but the density, temperature and electron
fraction were not interpolated between the grid points of the Monte Carlo
simulations. 
\par
\begin{figure}
\epsfxsize=8.8cm  \epsffile{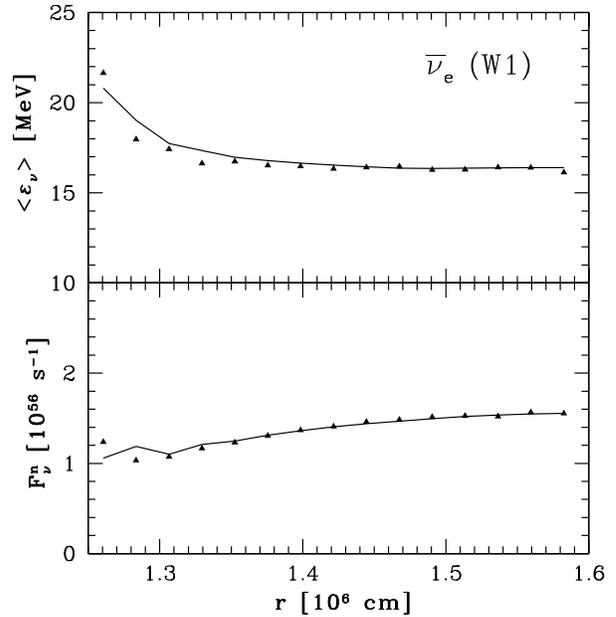} 
\caption[]{The radial evolution of the average energy of the flux 
(upper panel) and
the number flux (lower panel) of the electron-type anti-neutrino for
model~CS where 15 radial mesh points are used and the background
model is W1.  The filled triangles show the results of the Monte Carlo
simulation, while the solid lines represent the results of the Boltzmann 
simulation.}
\label{fig_15}

%\picplace{1cm}
%\figurenum{15}
%\caption{The radial evolution of the average energy (upper panel) and
%the number flux (lower panel) of the electron-type anti-neutrino for
%model~CS where 15 spatial mesh points are used and the background
%model is W1.  The solid triangles show the result of the Monte Carlo
%simulation, while the solid lines represent the result of the Boltzmann 
%simulation.
%\label{fig_15}}
\end{figure}

In Fig.~\ref{fig_15} we show the radial evolution  
of the average energy as defined in Eq.~(\ref{eqn:emin})
and that of the number flux given by Eq.~(\ref{eqn:nflx}) 
for model~CS, to be compared with the corresponding result for model ST2
in Fig.~\ref{fig_13}. It is clear that the agreement between the Monte Carlo
and the Boltzmann results for both quantities is good. 
We note also that the angular distribution as well as the 
energy spectrum are hardly affected by this change of the spatial resolution. 
Model~NI agrees with the Monte Carlo data nearly
perfectly (except for the problems with the angular distribution
discussed in Sect.~\ref{fefac}) after averaging over spatial mesh
zones in accordance to the way the Monte Carlo data represent the
transport result. Since the Boltzmann results do not change with the
number of radial grid points, we conclude that the quality of the 
numerical solutions is not degraded very much for simulations
with a decreased spatial resolution.

\par
Minor oscillations of the number flux near the inner boundary can be 
seen in Fig.~\ref{fig_13}. This problem results from the fact that one
cannot consistently specify the distribution of neutrinos which leave the 
computational volume at the inner boundary.
While this distribution should be determined from the 
transport result just above the inner boundary,  
the Boltzmann solver, however, requires an ad hoc specification in order  
to calculate the flux at the inner boundary. This 
leads inevitably to an inconsistency of the flux in the innermost
zone and thus to the observed oscillations. In fact, 
when an inhomogeneous spatial mesh was used in model~NU, in which the 
innermost grid zone was five times smaller than in the standard 
models, the oscillations were diminished as well. 
\par
\begin{figure}
\epsfxsize=8.8cm  \epsffile{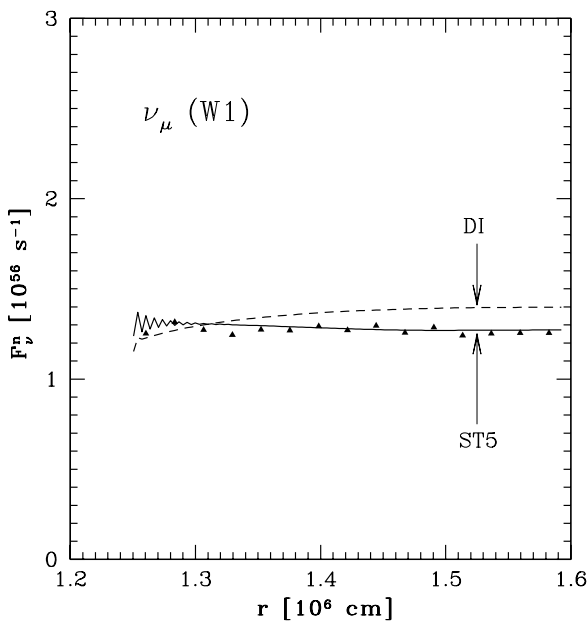} 
\caption[]{The number fluxes of the muon-type neutrino for 
model~DI (dashed line) and the corresponding standard model~ST4 (solid 
line). The triangles are the result of the Monte Carlo simulation.}
\label{fig_16}

%\picplace{1cm}
%\figurenum{16}
%\caption{The number fluxes of the muon-type neutrino for 
%model~DI (dashed line) and the corresponding standard model~ST4 (solid 
%line). The triangles are the results of the Monte Carlo simulation.
%\label{fig_16}}
\end{figure}

 
Finally, we illustrate possible errors which are associated with the treatment
of the finite differencing of the spatial advection term in the Boltzmann 
solver. In the radial advection term a linear average of the centered 
difference and of the upwind difference is used with a weight factor that 
changes according to the ratio of the neutrino mean free path to some chosen
length scale. Mezzacappa~et~al.~(\cite{me93a},\cite{me93b},\cite{me93c}) 
took the ratio of the mean free path 
to the local mesh width in order to construct the weighting. However, we 
found that this does not work well if the mesh width becomes of the same 
order as the mean free path but is much smaller than the scale height 
of the background. This is indeed the case in the inner optically
thick region of our standard models with 105 radial zones. 
In a more recent version of his code, Mezzacappa
(private communciation and~\cite{mez98}) defines the weighting factors
by refering them to the neutrinospheric radius.
In Fig.~\ref{fig_16} the dashed line shows the number flux of 
muon-type neutrinos for model~DI which used the prescription suggested 
by Mezzacappa~et~al.~(\cite{me93a},\cite{me93b},\cite{me93c}). 
The flux is slowly increasing with radius because the upwind
differencing is given too large a contribution in the optically 
thick region where the centered differencing should actually be
chosen. As demonstrated by the solid line in Fig.~\ref{fig_16}, 
the constancy of the flux, however, is recovered when 
the ratio of the mean free path to the distance up to the surface is
chosen instead of the ratio of the mean free path to the local mesh width.
Yet, this issue is probably not very important for realistic calculations 
of the entire neutron star, since the mesh width is usually not much smaller 
than the typical scale height of the matter distribution.

\par
To finish, we comment briefly on a last model in which  
we assumed that the nucleon scattering is taken isotropic to see to what
extent the result changes. No significant effect was found by
modifying the angular distribution of the dominant scattering reaction.

\section{Summary}

In this paper an extensive comparison was made between a newly developed 
Boltzmann neutrino transport code based on the discrete ordinate (S$_N$)
method as described by
Mezzacappa \& Bruenn~(\cite{me93a},\cite{me93b},\cite{me93c}),
and a Monte Carlo transport treatment (Janka~\cite{ja87}, \cite{ja91a}) 
by performing
time-dependent calculations of neutrino transport through realistic,
static profiles of protoneutron star atmospheres under the
assumption of spherical symmetry. In particular, the sensitivity
of the results of the Boltzmann solver to the employed numbers of 
radial, angular and energy grid points and to the treatment of the
radial advection terms was investigated. The flux factor and 
Eddington factor, which contain
information about the angular distribution of the neutrinos 
in the neutrino-decoupling region, were also compared with 
the approximate treatment of this regime by a multi-energy-group
flux-limited diffusion code (MGFLD).
\par 
The Boltzmann and Monte Carlo results showed excellent agreement
for observables such as the luminosity and the flux spectra which are
determined in those regions of the star where the neutrino-matter 
interactions are still very frequent and thus the neutrino
distributions are still nearly isotropic. Since the luminosity
and the spectra are essentially conserved farther out, the spatial 
evolutions as well as the surface values exhibit this agreement,
as long as the finite differencing of the Boltzmann solver is done
in a conservative way.
\par
Some problems, however, were observed concerning the description of
the angular distribution of the neutrinos by the Boltzmann results 
in the semitransparent and transparent regimes. 
Due to severe limitations of the number of angular grid points 
which can be used---typically only 6--10 angular bins between 
zero and 180 degrees are compatible with the steep increase of 
the requirements of computer time for better resolution---the 
Boltzmann code is not able to describe strong forward peaking of
the neutrino distributions very well, if a quadrature set is employed 
for the angular integration where the maximum angle cosine $\mu_{\rm max}$
is significantly less than unity. In this case the Boltzmann results
{\it underestimate} the anisotropy in the optically thin region outside 
the average ``neutrinosphere'', and the exact limits for the flux factor, 
$\left\langle \mu\right\rangle\to 1$, and for the Eddington factor,
$\left\langle \mu^2\right\rangle\to 1$, at large
distances away from the neutrino source cannot be satisfactorily
reproduced. This is in agreement with the trends also seen in recent
results of Messer~et~al.~(\cite{mm98}) and is exactly opposite to the
deficiencies of MGFLD which tends to {\it overestimate} the radial 
beaming of the radiation because flux limiters enforce 
the free-streaming limit when the optical depth 
of the stellar background becomes very low (Janka~\cite{ja91a}, 
\cite{ja92}).
\par
Since the dominant energy deposition rate by absorption of electron 
neutrinos and antineutrinos in the hot-bubble region of the
supernova core is inversely proportional to the flux factor
(see Eq.~(\ref{eq-1})), which means that the energy transfer from neutrinos
to the stellar plasma scales with the neutrino energy density (or number 
density) in the heating region, one cannot exclude that the 
Boltzmann solver may lead to an overestimation of the neutrino heating 
in disadvantageous situations, whereas MGFLD yields rates which are
definitely too low. For a set of typical post-core bounce situations,
Messer~et~al.~(\cite{mm98}), however, claim on grounds of numerical 
tests that the net neutrino heating is converged with S$_6$ and that
the differences between S$_6$ and S$_4$ are minor.
The problems may be more serious for neutrino reactions like
neutrino-antineutrino annihilation which are sensitive to both the
flux factor $\left\langle \mu\right\rangle$ and the Eddington 
factor $\left\langle \mu^2\right\rangle$ of the neutrinos 
(see Janka~\cite{ja91b}).
\par
The description of the angular neutrino distribution
by the Boltzmann solver and thus the agreement with the highly 
accurate Monte Carlo data
can be significantly improved by employing a variable angular mesh, 
even without increasing the total number of angular mesh points.
The positions of the angular grid points must be moved at each
time step and in every spatial zone such that they are clustering
in the forward direction in the optically thin regime.
\par
We found that the energy spectra can be well calculated with
about 15 energy mesh points. The fact that a reduction of the number
of spatial grid points from more than 100 to only 15 in the neutron 
star atmosphere did not change the quality of the transport results
means that the Boltzmann code can be reliably applied to realistic
simulations which involve the whole supernova core. Moreover, it
was demonstrated that the details of the interpolation between
centered differencing and upwind differencing in the spatial 
advection term can affect the accuracy of the transport results.
\par 
The excellent overall agreement of the results obtained with  
the Boltzmann code and the Monte Carlo method confirms the reliability 
of both of them. Good performance of the S$_N$ method for solving the
Boltzmann equation of neutrino transport has been found before by
Mezzacappa \& Bruenn~(\cite{me93a},\cite{me93b},\cite{me93c}) and
Messer~et~al.~(\cite{mm98}) even for realistic dynamic situations.
We hope that the work described here also helps to
reveal possible deficiencies and weaknesses and thus will serve
for further improvements of the numerical treatment
of neutrino transport in supernovae and protoneutron stars.

\begin{acknowledgements}
Encouraging discussions and helpful suggestions by E. M\"uller are
acknowledged. We are grateful to A. Mezzacappa for critical comments
on a first version of the paper and for updating us with the most 
recent improvements of his code. This work was partially
supported by the Japanese Society for the Promotion of Science (JSPS),
Postdoctoral Fellowships for Research Abroad, and by the 
Supercomputer Projects (No.97-22 and No.98-35) of the High Energy 
Accelerator Research Organization (KEK). H.-T.J. was supported, in part,
by the Deutsche Forschungsgemeinschaft under grant No.\ SFB-375.
The numerical calculations
were mainly done on the supercomputers of KEK. 
\end{acknowledgements}


\begin{thebibliography}{}
%\bibitem[1989]{ar89}   Arnett, W. D., Bahcall, J. N., 
%                       Kirshner, R. P. and Woosley, S. E.  1989, 
%                       ARA\&A, 27, 629 
%\bibitem[1991]{ba91}   Baron, E. and Cooperstein, J. 
%                       1991, Supernovae : The Tenth Santa Cruz 
%                       Summer Workshop in Astronomy
%                       and Astrophysics, edited by Woosley, S. E., 
%                      (Springer-Verlag, New York, 1991), p342
%\bibitem[1990]{bet90}  Bethe, H. A.  1990,  Rev. Mod. Phys., 62, 801
\bibitem[1985]{bw85}   Bethe, H. A. and Wilson. J. R. 1985, ApJ, 295, 14
\bibitem[1987]{bi87}   Bionta, R. M. et al. 1987, Phys. Rev. Lett., 58, 1494
\bibitem[1982]{bw82}   Bowers, R. L. and Wilson, J. R.  1982, ApJS, 50, 115 
%\bibitem[1994]{bb94}   Brown, G. E. and Bethe, H. A. 1994, ApJ, 423, 659
\bibitem[1985]{br85}   Bruenn, S. W. 1985, ApJS, 58, 771
%\bibitem[1986]{br86}   Bruenn, S. W. 1986, ApJS, 62, 331
%\bibitem[1989]{br89}   Bruenn, S. W. 1989, ApJ, 341, 385
%\bibitem[1993]{br93}   Bruenn, S. W. 1993, Proceedings of the
%                       First Symposium on ``Nuclear Physics 
%                       in the Universe'', held in Oak Ridge, 
%                       24 - 26 September 1992, edited by 
%                       Guidry, M. W. and Strayer, M. R., 
%                      (Inst. of Physics Publ., Bristol, 1993), p31
\bibitem[1987]{bu87}   Burrows, A.  1987, ApJ, 318, 57
\bibitem[1997]{bu97}   Burrows, A.  1997, to be published in the
                       proceedings of the 18'th Texas Symposium 
                       on Relativistic Astrophysics, 
                       held in Chicago, 15 - 20 December 1996, 
                       edited by Olinto, A., Frieman, J. and Schramm, D., 
                      (World Scientific Press, Singapore, 1997)
\bibitem[1993]{bg93}   Burrows, A. and Goshy, J. 1993, ApJ, 416, 75
\bibitem[1995]{bu95}   Burrows, A., Hayes, J. and Fryxell, B. A.  
                       1995, ApJ, 450, 830
\bibitem[1998a]{bs98a} Burrows, A., Sawyer, R. F. 1998a, 
                       Phys. Rev., C58, 554
\bibitem[b]{bs98b}     Burrows, A., Sawyer, R. F. 1998b, 
                       submitted to Phys. Rev. Lett.
\bibitem[1994]{cb94}   Cernohorsky, J. and Bludman, S. A. 1994, 
                       ApJ, 433, 250
\bibitem[1992]{dg92}   Dgani, R. and Janka, H.-Th.  1992, A\&A, 256, 428
\bibitem[1997]{hr97}   Hannestad, S. and Raffelt, G. 1997, 
                       to appear in ApJ
%\bibitem[1992]{he92}   Herant, M., Benz, W. and Colgate, S. A.  1992, 
%                       ApJ, 395, 642
\bibitem[1994]{he94}   Herant, M., Benz, W., Hix, J., Freyer, C. and 
                       Colgate, S. A.  1994, ApJ, 435, 339
%\bibitem[1987]{hl87}   Hillebrandt, W. 1987, High Energy Phenomena 
%                       Around Collapsed Stars, edited by Pacini, F., 
%                      (D. Reidel Publishing Company, Dordrecht, 1987), p73
\bibitem[1985]{hw85}   Hillebrandt, W. and Wolff, R. G.  1985, 
                       Nucleosynthesis : Challenges and New Developments, 
                       edited by Arnett, W. D. and Truran, J. W., 
                      (University of Chicago press, Chicago, 1985), p131
\bibitem[1987]{hi87}   Hirata, K. et al. 1987, Phys. Rev. Lett., 58, 1490
\bibitem[1991]{ho91}   Horowitz, C. J. and Wehrberger, K.  1991, 
                       Phys. Lett. B,  266, 236
\bibitem[1987]{ja87}   Janka, H.-Th.  1987,
                       Nuclear astrophysics; Proceedings of the
                       Workshop, Tegernsee, Germany, Apr.~21.--24., 1987,
                      (Springer-Verlag, Berlin and New York, 1987), p319
\bibitem[1991a]{ja91a} Janka, H.-Th.  1991a, Ph.D. thesis, 
                       Technische Universit\"{a}t M\"{u}nchen
\bibitem[1991b]{ja91b} Janka, H.-Th.  1991b, A\&A, 244, 378
\bibitem[1992]{ja92}   Janka, H.-Th.  1992, A\&A, 256, 452
\bibitem[1993]{ja93}   Janka, H.-Th.  1993, 
                       Frontier Objects in Astrophysics and Particle Physics,
                       Proc.~of the Vulcano Workshop 1992, Conf.~Proc.~Vol.~40,
                       edited by Giovannelli, F. and Mannocchi, G.,
                      (SIF, Bologna, 1993), p345
\bibitem[1992]{jd92}   Janka, H.-Th., Dgani, R. and 
                       van den Horn, L. J. 1992, A\&A, 265, 345 
\bibitem[1989a]{jh89a} Janka, H.-Th. and Hillebrandt, W.  1989a, 
                       A\&AS, 78, 375
\bibitem[b]{jh89b}     Janka, H.-Th. and Hillebrandt, W.  1989b,
                       A\&A, 224, 49
\bibitem[1998]{jk98}   Janka, H.-Th. and Keil, W. 1998, 
                       Supernovae and Cosmology, Proc. of a Colloquium
		       in Honor of Prof.~G.~Tammann on the Occasion 
                       of his 65th Birthday, 
		       Augst, Switzerland, Jun.~13., 1997,
		       edited by Labhardt, L., Binggeli, B. and Buser R.,
		       (Astronomisches Institut der Universit\"at Basel,
		       Basel, 1998) p7
\bibitem[1996]{jk96}   Janka, H.-Th., Keil, W., Raffelt, G. 
                       and Seckel, D. 1996, Phys. Rev. Lett., 76, 2621
\bibitem[1993]{jm93}   Janka, H.-Th. and M\"{u}ller, E.  1993,
                       Frontiers of Neutrino Astrophysics,
                       Proc.~of the International Symposium on 
                       Neutrino Astrophysics,
                       Takayama/Ka\-mioka, Japan, Oct.~19.--22., 1992,
                       edited by Suzuki, Y. and Nakamura, K.,
                      (Universal Academy Press, Tokyo, 1993) p203
%\bibitem[1995]{jm95}   Janka, H.-Th. and M\"{u}ller, E.  1995, 
%                       Phys. Rep., 256, 135 
\bibitem[1996]{jm96}   Janka, H.-Th. and M\"{u}ller, E.  1996, A\&A, 306, 167
\bibitem[1996]{ke96}   Keil, W., Janka, H.-Th. and M\"{u}ller, E.
                       1996, ApJ Lett., 473, L111
\bibitem[1995]{kj95}   Keil, W., Janka, H.-Th. and Raffelt, G. 1995,
                       Phys. Rev., D51, 6635
%\bibitem[1987]{la87}   Larsen, E. W., Morel, J. E. and
%                       Miller, W. F. Jr.  1987, J. Comp. Phys., 69, 283
%\bibitem[1992]{le92}   Leinson, L. B. 1992, Ap\&SS, 190, 271
%\bibitem[1981]{lp81}   Levermore, C. D. and Pomraning, G. C. 1981, 
%                       ApJ, 248, 321
%\bibitem[1978]{li78}   Lichtenstadt, I., Ron, A., Sack, N., 
%                       Wagschal, J. J. and Bludman, S. A.  1978, 
%                       ApJ, 226, 222
\bibitem[1998]{li98}   Lichtenstadt, I., Khokhlov, A. M. and Wheeler, J. C.
	               1998, submitted to ApJ
\bibitem[1988]{mw88}   Mayle, R. and Wilson, J. R. 1988, ApJ, 334, 909
\bibitem[1998]{mm98}   Messer, O. E. B., Mezzacappa, A., Bruenn, S. W.
                       and Guidry, M. W. 1998, submitted to ApJ,
                       astro-ph 9805276
\bibitem[1998]{mez98}  Mezzacappa, A. 1998, J. Computational and Applied 
		       Mathematics, in press 
\bibitem[1989]{me89}   Mezzacappa, A. and Matzner, R. A.  1989, ApJ, 343, 853 
\bibitem[1993a]{me93a} Mezzacappa, A. and Bruenn, S. W.  1993a, ApJ, 405, 637
\bibitem[b]{me93b}     Mezzacappa, A. and Bruenn, S. W.  1993b, ApJ, 405, 669
\bibitem[c]{me93c}     Mezzacappa, A. and Bruenn, S. W.  1993c, ApJ, 410, 740
\bibitem[1998]{me98}   Mezzacappa, A. et al. 1998, ApJ, 495, 911
%\bibitem[1984]{mi84}   Mihalas, D. and Mihalas, B. W.  1984, 
%                       Foundations of Radiation Hydrodynamics, 
%                      (Oxford Univ. Press, Oxford, 1984)
\bibitem[1993]{mw93}   Miller, D. S., Wilson, J. R. and Mayle, R. W. 1993, 
                       ApJ, 415, 278
\bibitem[1964]{msh64}  Misner, C. W. and Sharp, D. H. 1964, 
                       Phys. Rev, 136,  B571  
%\bibitem[1991]{mul91}  M\"{u}ller, E. 1991, Lecture Notes in Physics, 373, 97
\bibitem[1987]{my87}   Myra, E. S. et al. 1987, ApJ, 318, 744
\bibitem[1997]{pl97}   Prakash, M. et al. 1997, Phys. Rep., 280, 1
\bibitem[1995]{ra95}   Raffelt, G. G. and Seckel, D. 1995, 
                       Phys. Rev., D52, 1780
\bibitem[1996]{ra96}   Raffelt, G. G., Seckel, D. and Sigl, G.  1996, 
                       Phys. Rev., D54, 2784
\bibitem[1997]{rp97}   Reddy, S., Prakash, M. and Lattimer, J. M. 1997,
                       ApJ., 478, 689
\bibitem[1998a]{rp98a} Reddy, S., Prakash, M. and Lattimer, J. M. 1998a,
                       to appear in Proc. Second Oak Ridge Symposium on
                       Nuclear and Atomic and Nuclear Astrophysics
\bibitem[1998b]{rp98b} Reddy, S., Prakash, M. and Lattimer, J. M. 1998b,
                       Phys. Rev., D58, 1309
\bibitem[1989]{sa89}   Sawyer, R. F. 1989, Phys. Rev., C40, 865
\bibitem[1990]{sc90}   Schinder, P. J.  1990, ApJS, 74, 249
%\bibitem[1989]{sb89}   Schinder, P. J. and Bludman, S. A.  1989, ApJ, 346, 350
%\bibitem[1982a]{ss82a} Schinder, P. J. and Shapiro, S. L.  1982, ApJ, 259, 311
%\bibitem[1982b]{ss82b} Schinder, P. J. and Shapiro, S. L.  1982, ApJS, 50, 23
\bibitem[1994]{sh94}   Shimizu, T., Yamada, S. and Sato, K.  1994, 
                       ApJ Lett., 432, L119
\bibitem[1997]{sc97}   Smit, J. M., Cernohorsky, J. and 
                       Dullemond, C. P. 1997, A\&A, 325, 203
\bibitem[1998]{se98}   Sumiyoshi, K. and Ebisuzaki, T.  1998,
                       Parallel Computing, 24, 287
\bibitem[1990]{su90}   Suzuki, H.  1990, Ph.D. Thesis,
                       University of Tokyo
\bibitem[1993]{su93}   Suzuki, H.  1993, 
                       Frontiers of Neutrino Astrophysics,
                       Proc.~of the International Symposium on 
                       Neutrino Astrophysics,
                       Takayama/Ka\-mioka, Japan, Oct.~19.--22., 1992,
                       edited by Suzuki, Y. and Nakamura, K.,
                      (Universal Academy Press, Tokyo, 1993) p219
\bibitem[1994]{su94}   Suzuki, H.  1994, Physics and Astrophysics of 
                       Neutrinos, edited by Fukugita, M. and Suzuki, A.,
                      (Springer-Verlag, Tokyo, 1994), p763
\bibitem[1978]{tu78}   Tubbs, D. L.  1978, ApJS, 37, 287
\bibitem[1975]{ts75}   Tubbs, D. L. and Schramm, D. N.  1975, 
                       ApJ, 201, 467
\bibitem[1982]{wi82}   Wilson, J. R. 1982, Proc. Univ. Illinois
                       Meeting on Numerical Astrophysics
%\bibitem[1985]{wi85}   Wilson, J. R. 1985, Numerical Astrophysics, 
%                       edited by Centrella, J. M., LeBlanc, J. M. 
%                       and Bowers, R. L., 
%                      (Jones \& Bartlett, Boston, 1985), p422. 
\bibitem[1988]{wi88}   Wilson, J. R. 1988, private communication
\bibitem[1988]{wm88}   Wilson, J. R. and Mayle, R. W. 1988, 
                       Phys. Rep. 163, 63
\bibitem[1993]{wm93}   Wilson, J. R. and Mayle, R. W. 1993, 
                       Phys. Rep. 227, 97
\bibitem[1986]{wm86}   Wilson, J. R., Mayle, R. W., Woosley, S. E. and
                       Weaver, T. 1986, Ann. N. Y. Acad. Sci., 470, 267 
\bibitem[1997]{ya97}   Yamada, S.  1997, ApJ, 475, 720
%\bibitem[1977]{yu77}   Yueh, W. R. and Buchler, J. R. 1977, ApJ, 217, 565
\end{thebibliography}
\end{document}